# REGRESSION-WITH-RESIDUALS ESTIMATION OF MARGINAL EFFECTS: A METHOD OF ADJUSTING FOR TREATMENT-INDUCED CONFOUNDERS THAT MAY ALSO BE MODERATORS

**Geoffrey T. Wodtke**

University of Toronto

**Zahide Alaca**

University of Toronto

**Xiang Zhou**

Harvard University

**Corresponding Author:** Geoffrey T. Wodtke, Department of Sociology, University of Toronto, 725 Spadina Avenue, Toronto, ON M5S 2J4, Canada. Email: geoffrey.wodtke@utoronto.ca

**Acknowledgements:** This research was supported by an Insight Grant from the Social Sciences and Humanities Research Council of Canada (Grant No. 435-2018-0736).

**ABSTRACT**

Treatment-induced confounders complicate analyses of time-varying treatment effects and causal mediation. Conditioning on these variables naively to estimate marginal effects may inappropriately block causal pathways and may induce spurious associations between treatment and the outcome, leading to bias. Although several methods for estimating marginal effects avoid these complications, including inverse-probability-of-treatment-weighted (IPTW) estimation of marginal structural models (MSMs) as well as g- and regression-with-residuals (RWR) estimation of highly constrained structural nested mean models (SNMMs), each suffers from a set of nontrivial limitations. Specifically, IPTW estimation is inefficient, is difficult to use with continuous treatments or mediators, and may suffer from finite-sample bias, while g- and RWR estimation of highly constrained SNMMs for marginal effects are premised on the unrealistic assumption that there is no effect moderation. In this study, we adapt the method of RWR to estimate marginal effects with a set of moderately constrained SNMMs that accommodate several types of treatment-by-confounder and/or mediator-by-confounder interaction, thereby relaxing the assumption of no effect moderation. Through a series of simulation experiments and empirical examples, we show that this approach outperforms IPTW estimation of MSMs as well as both g- and RWR estimation of highly constrained SNMMs in which effect moderation is assumed away.

**INTRODUCTION**

In analyses of time-varying treatment effects or causal mediation, social scientists must often contend with the complications posed by treatment-induced confounders (e.g., Acharya et al. 2016; Elwert and Winship 2014; Wodtke et al. 2011). A treatment-induced confounder is a variable that is affected by a prior treatment and, in analyses of time-varying treatments, affects both selection into future treatment and the outcome, or, alternatively, in analyses of causal mediation, affects both the mediator and the outcome. For example, in analyses of whether living in a disadvantaged neighborhood throughout childhood and adolescence affects academic achievement (e.g., Sampson et al. 2008; Wodtke et al. 2011, 2016), parental income is likely affected by prior neighborhood conditions and also likely affects both future residential choices and child educational outcomes. Similarly, in analyses of whether family income mediates the effect of education on mental health (e.g., Cutler and Lleras-Muney 2006; Lee 2011), marital stability is likely affected by education and also confounds the effect of family income on mental health.

If left uncontrolled, treatment-induced confounders lead to bias in estimates of marginal effects, such as the cumulative treatment effect ($CTE$) in analyses of time-varying treatments or the controlled direct effect ($CDE$) in analyses of causal mediation. At the same time, adjusting naively for treatment-induced confounders by including them as predictors in a conventional regression model or matching on them via the propensity score also leads to bias. Specifically, conditioning on a treatment-induced confounder with conventional regression or matching methods leads to bias from over-control of intermediate pathways and endogenous selection (Elwert and Winship 2014; Robins et al. 2000; VanderWeele 2015). Alternative methods are therefore required when estimating marginal effects in the presence of treatment-induced confounders.

Fortunately, there are several methods that avoid the complications outlined previously and that are capable of consistently estimating marginal effects, even when adjustment is required for treatment-induced confounders. These include inverse-probability-of-treatment-weighted (IPTW) estimation of marginal structural models (MSMs; Robins et al. 1994, 2000), g-estimation of highly constrained structural nested mean models (SNMMs; Naimi et al. 2017; Vansteelandt 2009; Vansteelandt and Sjolander 2016), and regression-with-residuals (RWR) estimation of highly constrained SNMMs (Wodtke 2018).

Each of these methods, however, suffers from a set of nontrivial limitations. IPTW estimation is relatively inefficient, is difficult to use with continuous treatments or mediators, and may suffer from finite-sample bias when confounders strongly predict treatment and/or a mediator (Lunceford and Davidian 2004; Naimi et al. 2014; Robins et al. 1994). G- and RWR estimation of highly constrained SNMMs for marginal effects avoid the limitations of IPTW estimation, but they are premised on the strong assumption of no effect moderation (e.g., Vansteelandt 2009; Wodtke 2018), which is unrealistic in most social science applications. If, for example, a treatment-induced confounder also moderates the effect of a future treatment, or mediator, on the outcome, then these methods suffer from model misspecification bias. Because effect moderation is ubiquitous in the social sciences (Morgan and Winship 2015; Xie 2007), this assumption may limit the utility of these methods in practice.

In this study, we adapt the method of RWR to estimate a set of moderately constrained SNMMs for marginal effects that accommodate several types of treatment-by-confounder and/or mediator-by-confounder interaction. Briefly, RWR estimation of marginal effects in a moderately constrained SNMM proceeds in two stages. First, the confounders at each time point are regressed on all prior variables and then residualized. Second, the outcome is regressed on all prior variables, including a set of treatment-by-confounder and/or mediator-by-confounder interaction terms, with the residuals from the first stage substituted for the untransformed confounders both as "main effects" and as part of the interaction terms. Our adaptation differs from previous implementations of RWR (e.g., Almirall et al. 2010; Wodtke and Almirall 2017; Wodtke 2018) by additionally including the residualized confounders in interaction terms with treatment and/or a mediator, which accommodates several types of effect moderation while neatly isolating the marginal effects of interest in a single, possibly vector-valued, parameter.

Under the assumptions of sequential ignorability and no model misspecification, the proposed method is consistent for marginal effects, like the $CTE$ or $CDE$, even in the presence of treatment-induced confounders. It avoids the biases that arise with naive adjustments for treatment-induced confounders because the residualized confounders are purged of their association with prior treatment and thus including them in a regression model for the outcome is unproblematic. In addition, because it does not involve weighting by a function of the conditional probability of treatment and/or a mediator, the proposed method avoids the limitations associated with IPTW estimation. Finally, because it accommodates several types of treatment-by-confounder and/or mediator-by-confounder interaction, it also mitigates the limitations associated with both g- and RWR estimation of marginal effects using a highly constrained SNMM in which effect moderation is assumed away entirely.

In the sections that follow, we begin by considering the problem of estimating marginal effects for a time-varying treatment, such as the $CTE$. First, we formally define the effects of interest in the time-varying setting, explain when they are identified from observed data, and illustrate the problems that afflict conventional estimation methods in the presence of treatment-induced confounding. Second, we present an SNMM for the conditional, rather than marginal, effects of treatment, but we then show how these conditional effects can be additively decomposed into a set of functions that capture the marginal effects of interest and another set of functions that capture effect moderation. Third, we show how to appropriately parameterize these functions and adapt the method of RWR to estimate marginal effects with an SNMM under this alternative parameterization. Next, we build on this discussion by briefly considering analyses of causal mediation and the problem of estimating the $CDE$, which we show can be accomplished with the same methods used for estimating marginal effects in the time-varying setting. Finally, with a series of simulation experiments and empirical examples, we illustrate several applications of our proposed method and show that it outperforms other common approaches.

## NOTATION, ESTIMANDS, AND IDENTIFICATION

In this section, we formally define the marginal effects of interest and explain when they can be identified from observed data, drawing heavily on the potential outcomes framework (Holland 1986; Rubin 1974) and directed acyclic graphs (DAGs; Pearl 2009) throughout. For expositional clarity, we focus on a simplified example with a binary treatment measured at two time points, a binary confounder measured at two time points, and a continuous outcome measured at the end of follow-up, although these methods can be easily adapted for more complex analyses.

First, let $a_t = 1$ denote exposure to treatment, and $a_t = 0$ denote the absence of treatment, at time $t \in \{1,2\}$. Second, let $Y_i(a_1, a_2)$ denote the potential outcome for subject $i$ had she previously been exposed to the treatment sequence $\{a_1, a_2\}$. For example, $Y_i(0,0)$ is the potential outcome for subject $i$ had she never received treatment, $Y_i(1,0)$ is her outcome had she received treatment only at time $t = 1$, and so on. In this framework, each subject is conceived to have a potential outcome corresponding to each of the four possible treatment sequences, but only the single potential outcome corresponding to the treatment sequence actually received is ever observed in reality, and the others are so-called "counterfactuals." Third, let $C_{i1}$ denote the confounder for subject $i$ measured just prior to treatment at time $t = 1$, and let $C_{i2}(a_1)$ denote the confounder for subject $i$ measured just before treatment at time $t = 2$, which is indexed by $a_1$ as a potential outcome to reflect that it is affected by prior treatment. In other words, $C_{i2}(a_1)$ is a treatment-induced confounder. Finally, let the set $\{C_{i1}, A_{i1}, C_{i2}, A_{i2}, Y_i\}$ denote the observed data in temporal order.

In general, marginal effects are contrasts between different potential outcomes averaged over a population of individuals. More specifically, they give the average difference in the end-of-study outcome had everyone in the target population received one rather than another treatment sequence. With two time points, several different marginal effects may be of interest. The first is the distal treatment effect, or $DTE$, which can be formally defined as

$$DTE(a_1) = E\big(Y_i(a_1, 0) - Y_i(0,0)\big). \quad (1)$$

It gives the average effect of receiving treatment only at time $t = 1$ rather than never receiving treatment. The second is the proximal treatment effect, or $PTE$, which can be formally defined as

$$PTE(a_1, a_2) = E\big(Y_i(a_1, a_2) - Y_i(a_1, 0)\big). \quad (2)$$

When $a_1 = 0$, it gives the average effect of receiving treatment only at time $t = 2$ rather than never receiving treatment, and when $a_1 = 1$, it gives the average effect of always receiving treatment rather than receiving treatment only at time $t = 1$. The third is the cumulative treatment effect, or $CTE$. This effect is equal to the sum of $DTE(1)$ and $PTE(1,1)$,

$$CTE = DTE(1) + PTE(1,1) =$$

$$E\big(Y_i(1,0) - Y_i(0,0)\big) + E\big(Y_i(1,1) - Y_i(1,0)\big) = E\big(Y_i(1,1) - Y_i(0,0)\big), \quad (3)$$

which gives the average effect of being always versus never treated. Finally, the last is the interaction effect, or $INE$. This effect can be formally defined as

$$INE = PTE(1,1) - PTE(0,1) = E\big(Y_i(1,1) - Y_i(1,0)\big) - E\big(Y_i(0,1) - Y_i(0,0)\big), \quad (4)$$

which describes how the effect of receiving treatment at time $t = 2$ differs depending on whether an individual had previously received treatment at time $t = 1$.

All of these effects can be identified from the observed data under the assumptions of consistency, positivity, and sequential ignorability (Robins et al. 1994, 2000). The consistency assumption requires that the observed outcome $Y_i$ be equal to $Y_i(a_1, a_2)$ whenever $A_{i1} = a_1$ and $A_{i2} = a_2$. The positivity assumption requires that there not be any subgroups within the target population that are treated or untreated with certainty. The sequential ignorability assumption requires that the potential outcomes are independent of treatment at each time point conditional on the observed past. Formally, this assumption can be expressed as

$$Y_i(a_1, a_2) \perp A_{i1} | C_{i1} \; \forall \; (a_1, a_2) \text{ and } Y_i(a_1, a_2) \perp A_{i2} | C_{i1}, A_{i1}, C_{i2} \; \forall \; (a_1, a_2), \quad (5)$$

where $\perp$ denotes statistical independence. It is satisfied when there are not any unobserved variables that directly affect both selection into treatment at each time point and the outcome.

Figure 1 presents a DAG illustrating a set of causal relationships between the variables outlined previously in which the sequential ignorability assumption is satisfied. It shows that both treatments, $A_{i1}$ and $A_{i2}$, directly affect the outcome, $Y_i$, and that $A_{i1}$ also indirectly affects the outcome through $C_{i2}$. In addition, it shows that $C_{i1}$ confounds the effect of $A_{i1}$ on $Y_i$ and that $C_{i2}$ confounds the effect of $A_{i2}$ on $Y_i$. Treatment assignment is sequentially ignorable in this figure because treatment at each time point is not directly affected by any unobserved variables;

rather, the only unobserved variables, denoted by $U_{i1}$ and $U_{i2}$, directly affect the observed confounders and the outcome but not either treatment. The marginal effects outlined previously can be consistently estimated from the observed data by appropriately adjusting for all variables that directly affect both treatment and the outcome—in this case, $C_{i1}$ and $C_{i2}$.

**THE PROBLEM OF TREATMENT-INDUCED CONFOUNDING**

Because $C_{i2}$ is affected by $A_{i1}$ and confounds the effect of $A_{i2}$ on $Y_i$, it is a treatment-induced confounder. Treatment-induced confounders pose several challenges for estimating marginal effects of a time-varying treatment. In particular, conventional methods of covariate adjustment, including conditioning, stratifying, or matching directly on a treatment-induced confounder, lead to several types of bias, even when the effects of interest are identified under sequential ignorability. At the same time, failing to appropriately adjust for a treatment-induced confounder also leads to bias. Thus, treatment-induced confounders seemingly present a "damned if you do and damned if you don't" dilemma with regard to covariate adjustment.

To appreciate this, first consider the causal graph in Figure 2, and recall that a path in a DAG is "blocked" when it contains (a) an outcome of two or more variables, known as a collider, that has not been conditioned upon or (b) a non-collider that has been conditioned upon; otherwise, it is "unblocked" (Pearl 2009). Figure 2 shows that conditioning naively on the treatment-induced confounder, $C_{i2}$, blocks the causal pathway, $A_{i1} \rightarrow C_{i2} \rightarrow Y_i$, emanating from treatment at time $t = 1$ to the outcome, which leads to bias from over-control of intermediate pathways. Next consider the stylized graph in Figure 3. This figure shows that conditioning naively on $C_{i2}$ also unblocks the non-causal pathway, $A_{i1} \rightarrow C_{i2} \leftarrow U_{i2} \rightarrow Y_i$, emanating from treatment at time $t = 1$ to the outcome, which leads to bias from so-called "endogenous selection" or "collider stratification" (Elwert and Winship 2014). Specifically, it shows that $C_{i2}$ is a collider of $A_{i1}$ and $U_{i2}$, and because $U_{i2}$ affects $Y_i$, conditioning on $C_{i2}$ induces a spurious association between treatment at time $t = 1$ and the outcome. Finally, consider the stylized graph in Figure 4. This figure shows that when $C_{i2}$ has not been conditioned upon, the non-causal pathways emanating from treatment at time $t = 2$ to the outcome, $A_{i2} \leftarrow C_{i2} \rightarrow Y_i$ and $A_{i2} \leftarrow C_{i2} \leftarrow U_{i2} \rightarrow Y_i$, remain unblocked, which leads to bias from uncontrolled confounding. Thus, conventional methods of covariate adjustment inevitably lead to bias in estimates of marginal effects when there is treatment-induced confounding, and alternative methods are required.

## RWR FOR THE MARGINAL EFFECTS OF A TIME-VARYING TREATMENT

An SNMM is a model for the conditional, rather than marginal, effects of a time-varying treatment given the confounders (Almirall et al. 2010; Robins 1994; Robins et al. 2007; Wodtke and Almirall 2017). In this section, we show that conditional effects modeled with an SNMM can be additively decomposed into a set of functions that capture the marginal effects of interest and another set of functions that capture effect moderation. We then show how to appropriately parameterize these functions and adapt the method of RWR to consistently estimate them under the identification assumptions outlined previously and under the assumption of a correctly specified SNMM.

An SNMM is based on the following decomposition of the conditional mean of the potential outcomes given the confounders into a set of conditional treatment effects and a set of so-called "nuisance" associations:

$$E\big(Y_i(a_1, a_2)\big|C_{i1}, C_{i2}(a_1)\big) = \beta_{00} + \varepsilon_1(C_{i1}) + \mu_1(C_{i1}, a_1) + \varepsilon_2\big(C_{i1}, a_1, C_{i2}(a_1)\big) +$$

$$\mu_2\big(C_{i1}, a_1, C_{i2}(a_1), a_2\big), \quad (6)$$

where $\beta_{00} = E\big(Y_i(0,0)\big)$ is the marginal mean of the potential outcomes under no treatment;

$\varepsilon_1(C_{i1}) = \big[E(Y_i(0,0)|C_{i1}) - E\big(Y_i(0,0)\big)\big]$ is a nuisance association that captures the relationship

between the confounder at time $t = 1$ and the outcome under no treatment; $\mu_1(C_{i1}, a_1) =$

$E(Y_i(a_1, 0) - Y_i(0,0)|C_{i1})$ is a causal function that captures the conditional effects of treatment

at time $t = 1$ given $C_{i1}$; $\varepsilon_2\big(C_{i1}, a_1, C_{i2}(a_1)\big) = \big[E\big(Y_i(a_1, 0)\big|C_{i1}, C_{i2}(a_1)\big) - E(Y_i(a_1, 0)|C_{i1})\big]$ is

another nuisance association that captures the relationship between the confounder at time $t = 2$

and the outcome under treatment sequence $\{a_1, 0\}$; and $\mu_2\big(C_{i1}, a_1, C_{i2}(a_1), a_2\big) =$

$E\big(Y_i(a_1, a_2) - Y_i(a_1, 0)\big|C_{i1}, C_{i2}(a_1)\big)$ is another causal function that captures the conditional

effects of treatment at time $t = 2$ given both prior confounders. The functions $\varepsilon_1(C_{i1})$ and

$\varepsilon_2\big(C_{i1}, a_1, C_{i2}(a_1)\big)$ are called "nuisance" associations because they do not contain any

information about the causal effects of treatment (Wodtke and Almirall 2017).

The first causal function, $\mu_1(C_{i1}, a_1)$, can be further decomposed into a marginal effect of

interest and a term that captures effect moderation as follows:

$$\mu_1(C_{i1}, a_1) = \mu_{11}(a_1) + \mu_{12}(C_{i1}, a_1), \quad (7)$$

where $\mu_{11}(a_1) = E\big(Y_i(a_1, 0) - Y_i(0,0)\big)$ is equal to the $DTE(a_1)$ and $\mu_{12}(C_{i1}, a_1) =$

$\big[E(Y_i(a_1, 0) - Y_i(0,0)|C_{i1}) - E\big(Y_i(a_1, 0) - Y_i(0,0)\big)\big]$ captures how the effect of treatment at

time $t = 1$ differs across levels of $C_{i1}$.

Similarly, the second causal function, $\mu_2(C_{i1}, a_1, C_{i2}(a_1), a_2)$, can also be further

decomposed as follows:

$$\mu_2(C_{i1}, a_1, C_{i2}(a_1), a_2) = \mu_{21}(a_1, a_2) + \mu_{22}(C_{i1}, a_1, a_2) + \mu_{23}(C_{i1}, a_1, C_{i2}(a_1), a_2), \quad (8)$$

where $\mu_{21}(a_1, a_2) = E(Y_i(a_1, a_2) - Y_i(a_1, 0))$ is equal to another marginal effect of interest, the $PTE(a_1, a_2)$; $\mu_{22}(C_{i1}, a_1, a_2) = \left[ E(Y_i(a_1, a_2) - Y_i(a_1, 0)|C_{i1}) - E(Y_i(a_1, a_2) - Y_i(a_1, 0)) \right]$ captures how the effect of treatment at $t = 2$ differs across levels of $C_{i1}$; and $\mu_{23}(C_{i1}, a_1, C_{i2}(a_1), a_2) = \left[ E(Y_i(a_1, a_2) - Y_i(a_1, 0)|C_{i1}, C_{i2}(a_1)) - E(Y_i(a_1, a_2) - Y_i(a_1, 0)|C_{i1}) \right]$ captures how the effect of treatment at $t = 2$ differs across levels of $C_{i2}(a_1)$ within levels of $C_{i1}$.

Any parameterization of the marginal effects, $\mu_{11}(a_1)$ and $\mu_{21}(a_1, a_2)$, must satisfy the constraint that they are equal to zero when contemporaneous treatment is equal to zero. With a binary treatment, a saturated parameterization for $\mu_{11}(a_1)$ is

$$\mu_{11}(a_1) = \beta_{10}a_1, \quad (9)$$

and a saturated parameterization for $\mu_{21}(a_1, a_2)$ is

$$\mu_{21}(a_1, a_2) = (\beta_{20} + \beta_{21}a_1)a_2, \quad (10)$$

where $\beta_{10} = DTE(1)$, $\beta_{20} = PTE(0,1)$, and $\beta_{20} + \beta_{21} = PTE(1,1)$. In addition, note that $\beta_{10} + \beta_{20} + \beta_{21} = CTE$ and $\beta_{21} = INE$. Thus, all of the marginal effects defined previously are given by the parameter vector $\{\beta_{10}, \beta_{20}, \beta_{21}\}$.

Any parameterization of $\mu_{12}(C_{i1}, a_1)$ must satisfy the constraints that it is equal to zero when $a_1 = 0$ and that it has mean zero. With a treatment and confounder that are both binary, a saturated parameterization for this function is

$$\mu_{12}(C_{i1}, a_1) = \theta_{10}a_1 C_{i1}^\perp, \quad (11)$$

where $C_{i1}^\perp = \left( C_{i1} - E(C_{i1}) \right)$ is a residual transformation of $C_{i1}$ with respect to its marginal mean. This parameterization satisfies the zero mean constraint because $E(\theta_{10}a_1 C_{i1}^\perp) = \theta_{10}a_1 E(C_{i1}^\perp) = \theta_{10}a_1 E\left( \left( C_{i1} - E(C_{i1}) \right) \right) = \theta_{10}a_1\left( E(C_{i1}) - E(C_{i1}) \right) = 0$.

Similarly, any parameterization of $\mu_{22}(C_{i1}, a_1, a_2)$ must satisfy the constraints that it is equal to zero when $a_2 = 0$ and that it has mean zero. A saturated parameterization for this function is

$$\mu_{22}(C_{i1}, a_1, a_2) = (\theta_{20} + \theta_{21}a_1)a_2 C_{i1}^{\perp}, \quad (12)$$

which has mean zero because the expectation function is a linear operator and because $E(C_{i1}^{\perp}) = 0$, as above.

Finally, any parameterization of $\mu_{23}(C_{i1}, a_1, C_{i2}(a_1), a_2)$ must satisfy the constraints that it is equal to zero when $a_2 = 0$ and that it has mean zero conditional on $C_{i1}$. A saturated parameterization for this function is

$$\mu_{23}(C_{i1}, a_1, C_{i2}(a_1), a_2) = (\theta_{22} + \theta_{23}a_1 + (\theta_{24} + \theta_{25}a_1)C_{i1}^{\perp})a_2 C_{i2}^{\perp}(a_1), \quad (13)$$

where $C_{i2}^{\perp}(a_1) = \left(C_{i2}(a_1) - E(C_{i2}(a_1)|C_{i1})\right)$ is a residual transformation of $C_{i2}(a_1)$ with respect to its conditional mean given $C_{i1}$. This parameterization satisfies the zero mean constraint because the expectation function is a linear operator and because $E(C_{i2}^{\perp}(a_1)|C_{i1}) = E\left(\left(C_{i2}(a_1) - E(C_{i2}(a_1)|C_{i1})\right)|C_{i1}\right) = E(C_{i2}(a_1)|C_{i1}) - E(C_{i2}(a_1)|C_{i1}) = 0$. The parameter vector $\{\theta_{10}, \theta_{20}, \theta_{21}, \theta_{22}, \theta_{23}, \theta_{24}, \theta_{25}\}$ captures how the confounders moderate the effect of treatment at each time point.

The nuisance associations, $\varepsilon_1(C_{i1})$ and $\varepsilon_2(C_{i1}, a_1, C_{i2}(a_1))$, must also be parameterized under the constraint that they have mean zero given the past, which can be accomplished using the same residualized confounders as defined previously. Specifically, a saturated parameterization for the first nuisance association is

$$\varepsilon_1(C_{i1}) = \gamma_{10}C_{i1}^{\perp}, \quad (14)$$

and a saturated parameterization for the second nuisance association is

$$\varepsilon_2(C_{i1}, a_1, C_{i2}(a_1)) = (\gamma_{20} + \gamma_{21}a_1 + (\gamma_{22} + \gamma_{23}a_1)C_{i1}^{\perp})C_{i2}^{\perp}(a_1), \quad (15)$$

where the parameter vector $\{\gamma_{10}, \gamma_{20}, \gamma_{21}, \gamma_{22}, \gamma_{23}\}$ captures the associational (i.e., causal and possibly non-causal) effects of the confounders on the outcome.

Combining parametric expressions for the causal functions and nuisance associations yields the following saturated SNMM:

$$E\big(Y_i(a_1, a_2)|C_{i1}, C_{i2}(a_1)\big) = \beta_{00} + \gamma_{10}C_{i1}^{\perp} + \beta_{10}a_1 + \theta_{10}a_1C_{i1}^{\perp} + (\gamma_{20} + \gamma_{21}a_1 +$$

$$(\gamma_{22} + \gamma_{23}a_1)C_{i1}^{\perp})C_{i2}^{\perp}(a_1) + (\beta_{20} + \beta_{21}a_1)a_2 + (\theta_{20} + \theta_{21}a_1)a_2C_{i1}^{\perp} + (\theta_{22} + \theta_{23}a_1 +$$

$$(\theta_{24} + \theta_{25}a_1)C_{i1}^{\perp})a_2C_{i2}^{\perp}(a_1). \quad (16)$$

This model differs from that outlined in Almirall et al. (2010) and Wodtke and Almirall (2017) in that the residualized confounders are included not only in the nuisance associations but also as part of interaction terms in the causal functions. It also differs from the highly constrained SNMMs outlined in Vansteelandt and Sjolander (2016) and Wodtke (2018) in that effect moderation is not assumed to be absent but rather is explicitly modeled, or in other words, $\{\theta_{10}, \theta_{20}, \theta_{21}, \theta_{22}, \theta_{23}, \theta_{24}, \theta_{25}\}$ are free parameters that are not assumed to be zero.

An SNMM parameterized as above can be estimated using RWR, which proceeds in two stages. In the first stage, residual terms are estimated by centering $C_{i1}$ around its sample mean and by centering $C_{i2}$ around its estimated conditional mean given $C_{i1}$ and $A_{i1}$. Specifically, $\hat{C}_{i1}^{\perp} = C_{i1} - \hat{E}(C_{i1})$ and $\hat{C}_{i2}^{\perp} = C_{i2} - \hat{E}(C_{i2}|C_{i1}, A_{i1})$, where $\hat{E}(C_{i1}) = \frac{1}{n}\sum_i C_{i1}$ and $\hat{E}(C_{i2}|C_{i1}, A_{i1})$ is estimated from a generalized linear model with, for example, the logit or probit link when $C_{i2}$ is binary. Then, in the second stage, least squares estimates are computed for a linear regression of the outcome on prior treatments, the residualized confounders, and interactions involving the prior treatments and residualized confounders. This regression can be expressed as follows:

$$\hat{E}(Y_i|C_{i1}, A_{i1}, C_{i2}, A_{i2}) = \hat{\beta}_{00} + \hat{\gamma}_{10}\hat{C}_{i1}^{\perp} + \hat{\beta}_{10}A_{i1} + \hat{\theta}_{10}A_{i1}\hat{C}_{i1}^{\perp} + (\hat{\gamma}_{20} + \hat{\gamma}_{21}A_{i1} +$$

$$(\hat{\gamma}_{22} + \hat{\gamma}_{23}A_{i1})\hat{C}_{i1}^{\perp})\hat{C}_{i2}^{\perp} + (\hat{\beta}_{20} + \hat{\beta}_{21}A_{i1})A_{i2} + (\hat{\theta}_{20} + \hat{\theta}_{21}A_{i1})A_{i2}\hat{C}_{i1}^{\perp} + (\hat{\theta}_{22} + \hat{\theta}_{23}A_{i1} +$$

$$(\hat{\theta}_{24} + \hat{\theta}_{25}A_{i1})\hat{C}_{i1}^{\perp})A_{i2}\hat{C}_{i2}^{\perp}. \quad (17)$$

where different combinations of the estimated beta coefficients, $\{\hat{\beta}_{00}, \hat{\beta}_{10}, \hat{\beta}_{20}, \hat{\beta}_{21}\}$, are consistent for the marginal effects of interest under the identification assumptions outlined previously and under the assumption that the model is correctly specified, which is here assured by saturating it. This approach is nearly identical to conventional least squares regression except that the confounders at each time point are first residualized with respect to the observed past.

Figure 5 displays a stylized graph that illustrates the logic of RWR estimation. It shows that residualizing the confounders at each time point with respect to the observed past purges the treatment-induced confounder, $C_{i2}$, of its association with prior treatment, $A_{i1}$. As a result, including the residual transformation of $C_{i2}$ in a model for the outcome avoids bias due to over-control and endogenous selection. In addition, because RWR adjusts for observed confounding by conditioning on residual transformations of the confounders in an outcome regression rather than by re-weighting the data to appropriately balance the confounders across future treatments, it also avoids the limitations associated with IPTW estimation, such as the difficulty associated with constructing well-behaved weights for continuous treatments. Finally, by including the residualized confounders as part of interaction terms with treatment, RWR can accommodate effect moderation while neatly isolating the marginal effects of interest in a single parameter vector.

In practice, estimating a saturated SNMM is often impractical, or even impossible, either because the confounders or treatments are continuous or because there are a large number of time periods. In this situation, a set of parametric constraints must be imposed on the SNMM to facilitate estimation. For example, an analyst might consider excluding all higher-order interactions involving both of the confounders, in which case RWR estimation would proceed exactly as outlined previously except with the outcome regression in the second stage simplified as follows:

$$\hat{E}(Y_i|C_{i1}, A_{i1}, C_{i2}, A_{i2}) = \hat{\beta}_{00} + \hat{\gamma}_{10}\hat{C}_{i1}^{\perp} + \hat{\beta}_{10}A_{i1} + \hat{\theta}_{10}A_{i1}\hat{C}_{i1}^{\perp} + (\hat{\gamma}_{20} + \hat{\gamma}_{21}A_{i1})\hat{C}_{i2}^{\perp} +$$

$$(\hat{\beta}_{20} + \hat{\beta}_{21}A_{i1})A_{i2} + (\hat{\theta}_{20} + \hat{\theta}_{21}A_{i1})A_{i2}\hat{C}_{i1}^{\perp} + (\hat{\theta}_{22} + \hat{\theta}_{23}A_{i1})A_{i2}\hat{C}_{i2}^{\perp}. \quad (18)$$

Of course, many other constraints are possible, but recall that RWR requires a correctly specified model for the outcome. Thus, if any of these modeling constraints are inappropriate, then RWR is biased, even when the effects of interest are identified under sequential ignorability.

Additional modeling considerations are also required with RWR when there are multiple different confounders for which adjustment is necessary. First, all of the different confounders must be appropriately residualized in the first stage. This is accomplished by fitting a model for each confounder at each time point using all prior variables as predictors, and then extracting its residuals. Second, all of the residualized confounders must be included in the second-stage regression for the outcome, which may now involve additional interaction terms between treatment and the residualized confounders.

When estimating marginal effects with RWR and multiple different confounders, the method can accommodate all types of treatment-by-confounder interaction except for higher-order (i.e., three-way and above) interactions involving treatment and two or more different confounders measured contemporaneously. In the presence of such higher-order interactions, the conditional effects of treatment cannot be conveniently decomposed and parameterized with residual terms. Thus, with multiple different confounders, RWR estimation of marginal effects is suitable for a moderately constrained SNMM in which some especially complex forms of effect moderation are assumed to be absent. Although somewhat limiting, this modeling constraint is still considerably weaker than that required of other methods for estimating marginal effects with an SNMM (e.g., Wodtke 2018).

In sum, RWR estimation of a moderately constrained SNMM for marginal effects is a relatively simple adaptation of conventional least squares regression. It proceeds as follows: first, the confounders at each time point are regressed on all prior variables and then residualized, and second, the outcome is regressed on prior treatments, the residualized confounders, and to accommodate effect moderation, an admissible set of interaction terms involving prior treatments and the residualized confounders. The proposed method can accommodate all types of effect moderation except for those involving higher-order interactions between treatment and two or more different confounders measured at the same point in time. Marginal effect estimates can be constructed from the coefficients on prior treatments and any treatment-by-treatment interaction terms, while the magnitude and pattern of effect moderation is given by the coefficients on the interaction terms involving treatment and the residualized confounders. RWR is consistent under the identification assumptions outlined previously along with the assumption of a correctly specified SNMM. Valid standard errors can be obtained using the nonparametric bootstrap (Almirall et al. 2014).

**RWR FOR MARGINAL EFFECTS IN ANALYSES OF CAUSAL MEDIATION**

In this section, we briefly demonstrate that the methods outlined previously can also be used to estimate marginal effects in analyses of causal mediation. To appreciate this, first let $d$ denote exposure to a binary treatment, and let $m$ denote a putative mediator that is also binary. In addition, let $Y_i(d, m)$ denote the potential outcome for subject $i$ had she previously been exposed to treatment $d$ and the mediator $m$. Finally, let $X_i$ denote a treatment-outcome confounder for subject $i$ measured at baseline, and let $Z_i(d)$ denote a post-treatment confounder of the mediator-outcome relationship, which is indexed as a potential outcome by $d$ to reflect that it is a treatment-induced confounder.

In analyses of causal mediation, several different marginal effects may be of interest, but researchers often focus on a quantity called the $CDE$, which measures the causal relationship between treatment and the outcome when the putative mediator is fixed at the same value for all individuals. This estimand is useful because it is identified under weaker assumptions than others that may be of interest in mediation analyses, such as natural direct and indirect effects, and because it helps to adjudicate between different explanations for why treatment affects the outcome (Joffe and Greene 2009; VanderWeele 2009, 2015; Vansteelandt 2009). The $CDE$ can be formally defined as

$$CDE(d, m) = E\big(Y_i(d, m) - Y_i(0, m)\big). \quad (19)$$

In words, this quantity represents the average effect of treatment on the outcome when the mediator is fixed at the value $m$ for all individuals.

The $CDE$ can be identified from the observed data, here denoted in temporal order by the set $\{X_i, D_i, Z_i, M_i, Y_i\}$, under the assumptions of consistency, positivity, and sequential ignorability (VanderWeele 2009, 2015). In this context, the sequential ignorability assumption can be formally expressed as

$$Y_i(d, m) \perp D_i | X_i \; \forall \; (d, m) \text{ and } Y_i(d, m) \perp M_i | X_i, D_i, Z_i \; \forall \; (d, m), \quad (20)$$

which is satisfied when there are no unobserved treatment-outcome or mediator-outcome confounders.

Figure 6 presents a DAG illustrating a set of causal relationships between the variables outlined previously in which the sequential ignorability assumption is satisfied. The figure shows that both treatment, $D_i$, and the mediator, $M_i$, directly affect the outcome, $Y_i$. It also shows that $X_i$ confounds the effect of $D_i$ on $Y_i$ and that $Z_i$ confounds the effect of $M_i$ on $Y_i$. Finally, it shows that $D_i$ indirectly affects the outcome through $Z_i$, the mediator-outcome confounder. The sequential ignorability assumption is satisfied because the only unobserved variables, denoted by $U_i$ and $L_i$, do not directly affect treatment or the mediator. The $CDE$ can therefore be estimated from the observed data without bias by appropriately adjusting for $X_i$ and $Z_i$. Note, however, that the DAG in Figure 6 is structurally equivalent to those discussed previously for the time-varying setting, even though it contains a different set of variables. This indicates that conventional methods of covariate adjustment are also biased when estimating the $CDE$ if there are treatment-induced confounders, like $Z_i$, for the effect of the mediator on the outcome.

While conventional methods are biased in the presence of treatment-induced confounders, the $CDE$ can still be consistently estimated using an SNMM and RWR (Zhou and Wodtke 2018). For example, consider the following moderately constrained SNMM for the joint effects of treatment and the mediator on the outcome,

$$E\big(Y_i(d,m)|X_i,Z_i(d)\big) = \beta_{00} + \gamma_{10}X_i^\perp + \beta_{10}d + \theta_{10}dX_i^\perp + (\gamma_{20} + \gamma_{21}d + \gamma_{22}X_i^\perp)Z_i^\perp(d) +$$

$$(\beta_{20} + \beta_{21}d)m + (\theta_{20} + \theta_{21}d)mX_i^\perp + (\theta_{22} + \theta_{23}d)mZ_i^\perp(d), \quad (21)$$

where $X_i^\perp = X_i - E(X_i)$, $Z_i^\perp(d) = Z_i(d) - E(Z_i(d)|X_i)$, and, for simplicity, higher-order interactions involving both confounders are excluded. With this model, the $CDE$ is given by

$$CDE(d,m) = (\beta_{10} + \beta_{21}m)d, \quad (22)$$

and any potential moderation of the treatment effect by the baseline confounder, $X_i$, is captured by $\theta_{10}$ while any potential moderation of the mediator effect by $X_i$ or the post-treatment confounder, $Z_i(d)$, is captured by $\{\theta_{20}, \theta_{21}, \theta_{22}, \theta_{23}\}$.

This model can be estimated with RWR by, first, centering $X_i$ around its sample mean and centering $Z_i$ around its estimated conditional mean given $X_i$ and $D_i$, and then second, fitting a least squares regression of the outcome on treatment, the mediator, the residualized confounders, and a set of interaction terms between treatment, the mediator, and the residualized confounders. Specifically, the second-stage regression for the outcome can be expressed as

$$\hat{E}(Y_i|X_i, D_i, Z_i, M_i) = \hat{\beta}_{00} + \hat{\gamma}_{10}\hat{X}_i^{\perp} + \hat{\beta}_{10}D_i + \hat{\theta}_{10}D_i\hat{X}_i^{\perp} + (\hat{\gamma}_{20} + \hat{\gamma}_{21}D_i + \hat{\gamma}_{22}\hat{X}_i^{\perp})\hat{Z}_i^{\perp} +$$

$$(\hat{\beta}_{20} + \hat{\beta}_{21}D_i)M_i + (\hat{\theta}_{20} + \hat{\theta}_{21}D_i)M_i\hat{X}_i^{\perp} + (\hat{\theta}_{22} + \hat{\theta}_{23}D_i)M_i\hat{Z}_i^{\perp}, \quad (23)$$

where $\hat{X}_i^{\perp}$ and $\hat{Z}_i^{\perp}$ are the estimated residuals from the first stage. An RWR estimate of the $CDE$ is given by

$$\widehat{CDE}(d, m) = (\hat{\beta}_{10} + \hat{\beta}_{21}m)d. \quad (24)$$

It is consistent under the identification assumptions outlined previously and the assumption of no model misspecification. Although these are strong assumptions, they are considerably weaker than those required when estimating the $CDE$ with a highly constrained SNMM in which effect moderation is assumed to be absent (e.g., Acharya et al. 2016; Vansteelandt 2009).

In sum, RWR estimation of an SNMM for the joint effects of a treatment and mediator on an outcome proceeds as follows. First, the baseline confounders are residualized by centering them around their sample means, and the post-treatment confounders are residualized by regressing them on treatment and the baseline confounders. Second, the outcome is regressed on treatment, the mediator, the residualized confounders, and an admissible set of interaction terms involving treatment, the mediator, and the residualized confounders. With multiple different confounders, RWR can accommodate all types of effect moderation except for higher-order interactions involving two or more different confounders measured at the same point in time. Estimates of the $CDE$ can be constructed from the coefficients on treatment and any treatment-by-mediator interactions, while the magnitude and pattern of effect moderation is given by the coefficients on interaction terms involving treatment, the mediator, and the residualized confounders. As in the time-varying setting, valid standard errors can be estimated with the nonparametric bootstrap.

**SIMULATION EXPERIMENTS**

We use a series of simulation experiments to evaluate the performance of RWR estimation for marginal effects relative to other methods. Specifically, we use 10,000 simulations of $n = 500$ to estimate the $CTE$ of a time-varying treatment measured at two time points. In each simulation, we generate an "unobserved" continuous variable $U_i$, an observed continuous time-varying confounder $\{C_{i1}, C_{i2}\}$, a binary time-varying treatment $\{A_{i1}, A_{i2}\}$, and a continuous end-of-study outcome, $Y_i$. In these simulations, $[U_i] \sim N\left(\mu_{U_i} = 0, \sigma_{U_i}^2 = 1\right)$; $[C_{i1}] \sim N\left(\mu_{C_{i1}} = 0, \sigma_{C_{i1}}^2 = 1\right)$;

$[C_{i2}|U_i, C_{i1}, A_{i1}] \sim N\left(\mu_{C_{i2}|U_i, C_{i1}, A_{i1}} = 0.5U_i + 0.5C_{i1} + 0.5A_{i1}, \sigma_{C_{i2}|U_i, C_{i1}, A_{i1}}^2 = 1\right)$;

$[A_{i1}|C_{i1}] \sim \text{Bernoulli}\left(p_{A_{i1}} = \Phi(\gamma C_{i1})\right)$; $[A_{i2}|C_{i1}, A_{i1}, C_{i2}] \sim \text{Bernoulli}\left(p_{A_{i2}|C_{i1}, A_{i1}, C_{i2}} = \Phi(\gamma C_{i1} + 0.5A_{i1} + \gamma C_{i2})\right)$; $[Y_i|U_i, C_{i1}, A_{i1}, C_{i2}, A_{i2}] \sim N\left(\mu_{Y_i|U_i, C_{i1}, A_{i1}, C_{i2}, A_{i2}} = 0.5U_i + \gamma\left(C_{i1} - \mu_{C_{i1}}\right) + A_{i1}\left(0.2 + \theta\left(C_{i1} - \mu_{C_{i1}}\right)\right) + \gamma\left(C_{i2} - \mu_{C_{i2}}\right) + A_{i2}\left(0.2 + 0.1A_{i1} + \theta\left(\left(C_{i1} - \mu_{C_{i1}}\right) + \left(C_{i2} - \mu_{C_{i2}}\right)\right)\right), \sigma_{Y_i|U_i, C_{i1}, A_{i1}, C_{i2}, A_{i2}}^2 = 1\right)$, where $\Phi$ is the standard normal cumulative distribution function, and $\gamma$ and $\theta$ are parameters used to modify, respectively, the

level of treatment-outcome confounding and the magnitude of treatment effect moderation in different simulations. In all simulations, the $CTE$ is identified and its true value is 0.5.

We compare the performance of RWR estimation of a moderately constrained SNMM for marginal effects (henceforth "RWR with interactions") to the performance of conventional least squares regression, IPTW estimation of an MSM, g-estimation of a highly constrained SNMM in which effect moderation is assumed to be absent, and RWR estimation of the same highly constrained SNMM (henceforth "RWR without interactions"). To compute conventional regression estimates, we fit by least squares a linear regression of the outcome on prior treatments, the observed confounders, and a treatment-by-treatment interaction. The estimated $CTE$ is then given by the sum of the coefficients on prior treatments and the interaction term.

To compute IPTW estimates (Robins et al. 1994, 2000), we fit a linear regression of the outcome on prior treatments and their interaction using weighted least squares, with weights equal to

$$w_i = \frac{P(A_{i1} = a_{i1})}{P(A_{i1} = a_{i1}|C_{i1})} \times \frac{P(A_{i2} = a_{i2}|A_{i1} = a_{i1})}{P(A_{i2} = a_{i2}|C_{i1}, A_{i1} = a_{i1}, C_{i2})}, \quad (25)$$

where $w_i$ is estimated from a series of probit models for the conditional probabilities in the numerator and denominator of the weight. At each time point, weighting by $w_i$ balances (in expectation) prior confounders across future treatments by giving more weight to subjects with confounder histories that are underrepresented in a treatment group and less weight to subjects with confounder histories that are overrepresented in a treatment group. The estimated $CTE$ is the sum of the coefficients on prior treatments and their interaction.

To compute g-estimates of marginal effects using a highly constrained SNMM without any effect moderation, we use the g-estimator proposed by Vansteelandt and Sjolander (2016). Specifically, we first fit a linear regression of the outcome on prior treatments and their interaction, estimated propensity scores for treatment at each time point, an interaction between treatment at time $t = 1$ and the estimated propensity score for treatment at time $t = 2$, and the observed confounders at both time points. The coefficients on treatment at time $t = 2$ and its interaction with treatment at time $t = 1$ from this model provide estimates of the proximal treatment effect. Then, we subtract the estimated proximal treatment effect from the outcome for each respondent and regress this transformed outcome on the treatment, propensity score, and the observed confounder at time $t = 1$, where the coefficient on treatment from this model provides an estimate of the distal treatment effect. The estimated $CTE$, then, is the sum of the distal and proximal treatment effects computed as above. Vansteelandt and Sjolander (2016) show that this estimator is asymptotically equivalent to the doubly robust g-estimator considered in Robins et al. (1992).

To compute estimates based on RWR without interactions, we first residualize the observed confounders at each time point by regressing them on all prior variables and then centering them around their estimated conditional means. Second, we regress the outcome on prior treatments and their interaction as well as all residualized confounders. The estimated $CTE$ is the sum of the coefficients on prior treatments and their interaction. Computing estimates based on RWR with interactions proceeds in almost exactly the same manner except that all two-way interactions between the treatments and residualized confounders are additionally included in the second-stage regression for the outcome. Part A of the Online Supplement presents the R code used to execute all of the simulations outlined previously.

We compare the performance of these methods in terms of their bias, standard deviation, and root mean squared error (RMSE) under different levels of treatment-outcome confounding and under different levels of effect moderation. Because treatment-induced confounding is present in all simulations, we expect conventional regression to perform poorly across all scenarios. Because IPTW estimation is relatively inefficient and susceptible to finite-sample bias when confounders strongly predict treatment, we expect its performance to suffer in simulations with high levels of treatment-outcome confounding. Because g- and RWR estimation of marginal effects using a highly constrained SNMM require that the confounders must not moderate the effects of treatment, we expect their performance to deteriorate in simulations with high levels of treatment effect moderation. Finally, because RWR with interactions accommodates this type of effect moderation, we expect it to perform well across all simulations.

Table 1 presents results from a first set of simulation experiments, wherein we varied the level of treatment-outcome confounding in the absence of effect moderation. Conventional regression is badly biased at all levels of confounding, as expected. IPTW estimation performs well at lower levels of confounding but suffers from finite-sample bias at higher levels and is relatively inefficient, also as expected. G- and both types of RWR estimation perform similarly in these simulations: they are all unbiased and achieve comparable efficiency gains relative to IPTW estimation.

Table 2 presents results from a second set of simulation experiments, wherein we varied the level of treatment effect moderation after setting the level of treatment-outcome confounding at a moderate-to-high level. As expected, both conventional regression and IPTW estimation perform poorly. Although IPTW estimation accommodates effect moderation, it still suffers from finite-sample bias due to the high level of confounding and is relatively inefficient. Also as expected, G-estimation and RWR estimation without interactions are increasingly biased as the magnitude of treatment effect moderation rises, whereas RWR with interactions is unbiased and achieves the lowest RMSE across all scenarios.

**THE CTE OF NEIGHBORHOOD POVERTY ON ACADEMIC ACHIEVEMENT**

The effects of neighborhood composition on child development have long concerned social scientists across several different disciplines (e.g., Chetty et al. 2016; Leventhal and Brooks-Gunn 2000; Sampson et al. 2008; Wodtke et al. 2011). To illustrate how the proposed method can be used with time-varying treatments, we estimate the $CTE$ of residence in a disadvantaged neighborhood throughout the early life course on adolescent math achievement using data from $n = 1{,}135$ individuals in the Panel Study of Income Dynamics – Child Development Supplement (PSID-CDS; Michigan Survey Research Center 2014).[1]

Previously, Wodtke (2018) estimated this effect with data from the PSID-CDS by fitting a conventional regression model using least squares, an MSM using IPTW, and a highly constrained SNMM without any effect moderation using RWR. In that analysis, RWR estimates indicated that long-term residence in a disadvantaged neighborhood has a severe negative effect on math achievement—an effect that is obscured by bias in conventional regression models and imprecisely captured by IPTW. These estimates, however, are premised on the strong and arguably unrealistic assumption of no effect moderation. We overcome this limitation by estimating the $CTE$ using RWR and a moderately constrained SNMM that includes all two-way treatment-by-confounder interactions. By additionally including these interactions, our reanalysis relaxes the strong assumption of no effect moderation required in Wodtke (2018).

---

[1] Some of the data used in this analysis are based on "sensitive data files" from the PSID-CDS, which were obtained under special contractual arrangements designed to protect the anonymity of respondents. These data are not available from the authors. Persons interested in obtaining sensitive data files from the PSID-CDS should contact psidhelp@isr.umich.edu. A set of replication files for this analysis, sans any sensitive data, is provided as part of the online supplementary material.

Specifically, we model the distal, proximal, and cumulative marginal effects of exposure to a disadvantaged neighborhood on adolescent math achievement using the following SNMM:

$$E\big(Y_i(a_1, a_2)|C_{i1}, C_{i2}(a_1)\big) = \beta_{00} + \gamma_{10}^T C_{i1}^\perp + \beta_{10}a_1 + \theta_{10}^T a_1 C_{i1}^\perp + \gamma_{20}^T C_{i2}^\perp(a_1) + \beta_{20}a_2 +$$

$$\theta_{20}^T a_2 C_{i1}^\perp + \theta_{21}^T a_2 C_{i2}^\perp(a_1), \quad (26)$$

where $\beta_{10} = DTE(1)$, $\beta_{20} = PTE(a_1, 1)$, and $\beta_{10} + \beta_{20} = CTE$. In this model, the outcome, $Y_i$, represents standardized scores on the Woodcock-Johnson applied problems test measured at the end of follow-up when individuals were age 13 to 17 (Woodcock and Johnson 1989). The time-varying treatment, $a_t$, is a standardized index of neighborhood disadvantage generated via a principal component analysis of multiple census tract characteristics, such as the poverty rate, unemployment rate, and median household income. Treatment is first measured during childhood when individuals were age 5 to 9 and then again during adolescence when they were age 11 to 15. Finally, $C_{i1}^\perp$ and $C_{i2}^\perp$ are vectors of residualized confounders. The first vector, $C_{i1}^\perp$, contains a set of time-invariant factors, such as race, gender, and birth cohort, as well as a set of time-varying characteristics, including equivalized family income, parental marital status, and lagged achievement test scores, all measured during early childhood. The second vector, $C_{i2}^\perp$, contains the same set of time-varying characteristics only now measured just before the onset of adolescence. Detailed information on sample and variable definitions can be found in Wodtke (2018).

The first row of Table 3 reports RWR estimates for the distal, proximal, and cumulative effects of living in a disadvantaged neighborhood based on the moderately constrained SNMM outlined previously. These are computed by, first, estimating residuals for each of the confounders. This involves centering the elements of $C_{i1}$ around their sample means and centering the elements of $C_{i2}$ around their estimated conditional means, which are computed from least squares regressions of $C_{i2}$ on the treatment and confounders measured earlier during childhood. Second, marginal effect estimates are computed by regressing the outcome on both treatments, the residualized confounders, and all two-way interactions between the treatments and residualized confounders. For comparative purposes, the second and third rows of Table 3 report RWR and g-estimates of marginal effects from a highly constrained SNMM in which all treatment-by-confounder interactions are excluded—that is, a model in which the confounders are assumed not to moderate the effects of treatment in any way. Part B of the Online Supplement presents the R code used to generate the results in this table.

All of the estimates in Table 3 indicate that the distal effect of childhood exposure to a disadvantaged neighborhood on adolescent math achievement is substantively small and fails to reach conventional significance thresholds, that the proximal effect of adolescent exposure is larger and statistically significant, and that the cumulative effect of sustained exposure is substantively large and highly significant. For example, according to these estimates, sustained exposure to a poor neighborhood one standard deviation above the national mean of the disadvantage index, rather than sustained exposure to a wealthy neighborhood one standard deviation below the national mean, is estimated to reduce adolescent math achievement by about $0.127 \times 2 = 0.254$ standard deviations. Although the results in Table 3 are highly consistent across the different methods employed, those generated via RWR estimation of a moderately constrained SNMM with all two-way treatment-by-confounder interactions are premised on much weaker assumptions about effect moderation.

**THE CDE OF EDUCATION ON MENTAL HEALTH**

A number of prior studies have investigated the causal relationship between education and mental health (e.g., Cutler and Lleras-Muney 2006; Heckman et al. 2018; Lee 2011), but the mechanisms underlying this causal link remain unclear. Education may improve mental health by providing access to higher economic status and greater resources, or it may affect mental health through other channels—for example, by providing greater access to health information and improving health behaviors. To illustrate the utility of RWR for analyses of causal mediation, we examine the CDE of college completion on mental health controlling for family income as a mediator. In this example, a comparison between the total effect and the CDE helps to adjudicate whether family economic status explains the mental health benefits of college completion.

We use data from $n = 2{,}719$ individuals in the National Longitudinal Survey of Youth 1979 (NLSY79) who were age 14-17 when they were first interviewed in 1979. First, we estimate the total effect of college completion using the following model:

$$E(Y_i(d)|X_i) = \beta_{00} + \gamma_{10}^T X_i^\perp + \beta_{10} d + \theta_{10}^T d X_i^\perp. \quad (27)$$

In this model, the outcome, $Y_i$, represents scores on the Center for Epidemiologic Studies Depression Scale (CES-D) when respondents were 40 years old. We standardize CES-D scores to have mean zero and unit variance, where a higher score implies greater depression. The treatment, $d$, is defined as completion of a four-year college degree by age 25. The residualized baseline confounders, $X_i^\perp$, include gender, race, Hispanic ethnicity, mother's years of schooling, father's presence in the home, number of siblings, urban residence, educational expectations, and percentile scores on the Armed Forces Qualification Test (AFQT). Under this specification, $\beta_{10}$ captures the total effect of college completion on depression at age 40.

We then model the joint effects of college completion and family income on depression using the following SNMM:

$$E\big(Y_i(d,m)|X_i, Z_i(d)\big) = \beta_{00} + \gamma_{10}^T X_i^\perp + \beta_{10}d + \theta_{10}^T d X_i^\perp + \gamma_{20}^T Z_i^\perp(d) + (\beta_{20} + \beta_{21}d)m +$$
$$\theta_{20}^T m X_i^\perp + \theta_{21}^T m Z_i^\perp(d), \quad (28)$$

where the mediator of interest, $m$, is the percentile rank of equivalized family income averaged over ages 36-40. The residualized post-treatment confounders, $Z_i^\perp$, include CES-D scores measured when respondents were 27-30 years old, the proportion of time married between 1990 and 1998, and the number of relationship transitions between 1990 and 1998. These variables capture mental health and family stability during young adulthood, which may be affected by treatment (college completion by age 25) and also affect both the mediator (family income between age 36 and 40) and the outcome (depression at age 40). In this model, the controlled direct effect is given by $CDE(d,m) = (\beta_{10} + \beta_{21}m)d$.

The first row of Table 4 reports estimates for the total and direct effects of college completion on depression computed using RWR with interactions. These estimates are obtained by, first, computing residuals for each of the baseline confounders $X_i$ and post-treatment confounders $Z_i$, which involves centering the elements of $X_i$ around their sample means and centering the elements of $Z_i$ around their estimated conditional means given the past. Then, the total effect and CDE are estimated by fitting the models described previously using these residual terms. In particular, the CDE is evaluated at $m = 0.5$, that is, when equivalized family income is fixed at the sample median. For comparative purposes, the second and third rows of Table 4 report RWR and g-estimates of total and direct effects from a highly constrained SNMM in which all treatment-by-confounder and mediator-by-confounder interactions are excluded—that is, from a model in which the confounders are assumed not to moderate the effects of treatment or the mediator on the outcome. Part C of the Online Supplement presents the R code used to generate the results in this table.

RWR with interactions yields a sizable and statistically significant total effect of education on mental health, where completing college is estimated to lower depression scores by 0.165 standard deviations. There is also evidence of mediation via family income based on these estimates. Specifically, the estimated CDE, when family income is fixed at its sample median, is much smaller in magnitude than the total effect, which suggests that a substantial portion of this effect operates through pathways involving family economic resources. By contrast, results based on g-estimation and RWR without interactions, which come from models that assume away effect moderation altogether, are somewhat different. Specifically, both g-estimation and RWR without interactions produce smaller estimates for the total effect along with estimates for the CDE that are closer in magnitude to the total effect. Taken together, these results suggest that naively assuming away effect moderation may induce bias in analyses of causal mediation and potentially lead to erroneous conclusions about the importance of particular mediating pathways.

## DISCUSSION AND CONCLUSIONS

In analyses of causal mediation and time-varying treatment effects, treatment-induced confounders often complicate efforts to estimate marginal effects. Several available methods avoid these complications, including IPTW estimation of MSMs as well as g- and RWR estimation of highly constrained SNMMs, but they are not without limitations. Specifically, the performance of IPTW is poor with continuous treatments and/or mediators, a high degree of confounding, and small samples, while both g- and RWR estimation of highly constrained SNMMs are biased for the marginal effects of interest when effect moderation is present. To overcome these limitations, we adapt the method of RWR to estimate marginal effects with a set of moderately constrained SNMMs that easily accommodate several types of effect moderation as well as continuous treatments and/or mediators. A series of simulation experiments indicate that the proposed method outperforms IPTW estimation of MSMs in general and that it outperforms both g- and RWR estimation of highly constrained SNMMs in the presence of effect moderation. Because the proposed method involves only simple and familiar computations, it is easily implemented with standard software, as we demonstrate across two empirical illustrations.

Nevertheless, despite its many advantages, RWR estimation of marginal effects is premised on a number of strong modeling assumptions. Specifically, it requires a correctly specified SNMM, which in turn requires that all of the causal functions and nuisance associations that compose this model are correctly specified. It also requires the absence of more complex forms of effect moderation involving two or more confounders measured contemporaneously, which complicates the decomposition and parameterization of the SNMM causal functions using residual terms. The assumption of a correctly specified SNMM may be reasonable with a relatively small number of confounders and time periods, but identifying a correct model may be challenging with high dimensional data.

In this situation, researchers might consider combining the methods proposed in this study with either IPTW or g-estimation to leverage their strengths while mitigating their weaknesses. For example, RWR could be used to adjust for a subset of the time-varying confounders that prove difficult to appropriately balance using IPTW. Then, a simplified SNMM involving only this subset of confounders and a more limited set of interaction terms could be fit by RWR to an appropriately weighted sample in which the remaining confounders have all been balanced. Alternatively, the confounders could first be residualized with respect to the observed past and then included in interaction terms with treatment and/or a mediator at each stage of the g-estimation procedure outlined by Vansteelandt and Sjolander (2016). This may provide some protection against bias due to misspecification of the nuisance associations in an SNMM, as g-estimation is doubly robust, while simultaneously accommodating several types of effect moderation in analyses of marginal effects.

In sum, RWR estimation of a moderately constrained SNMM for marginal effects provides an appealing alternative to IPTW estimation of MSMs and to both g- and RWR estimation of highly constrained SNMMs in which effect moderation is assumed away. The proposed method improves upon IPTW estimation in that it is more efficient, easy to use with continuous treatments and/or mediators, and avoids finite-sample bias when the magnitude of observed confounding is strong. It improves upon g- and RWR estimation of highly constrained SNMMs in that it can easily accommodate all but highly complex forms of effect moderation while still neatly isolating the marginal effects of interest in a single set of parameters. Although the proposed method is premised on a number of strong modeling assumptions, it can be integrated with IPTW or g-estimation in situations where these assumptions are questionable to enhance its robustness. Given their flexibility, efficiency, and ease of use, we expect moderately constrained SNMMs along with the associated method of RWR to be frequently used in future studies of causal mediation and time-varying treatment effects.

**FIGURES**

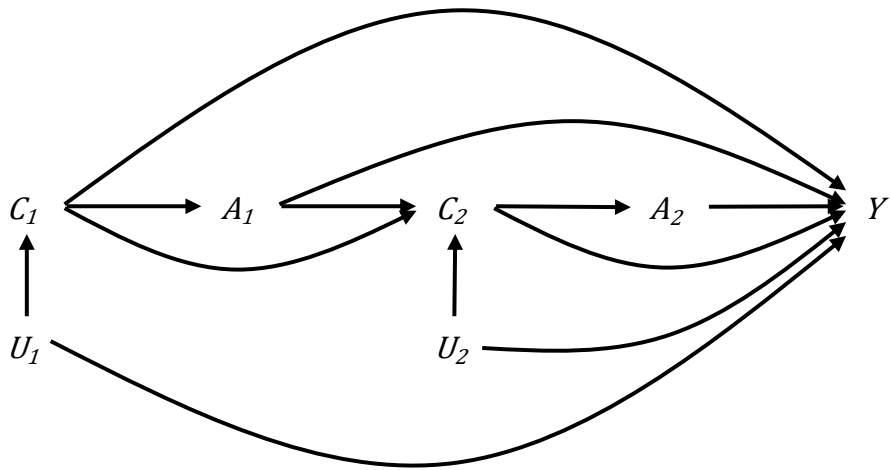

Figure 1. A directed acyclic graph illustrating a set of causal relationships between a time-varying treatment, a time-varying confounder, and an outcome

Notes: $A_t$ denotes a time-varying treatment, $C_t$ denotes an observed time-varying confounder, $U_t$ denotes an unobserved time-varying covariate, and $Y$ denotes an end-of-study outcome.

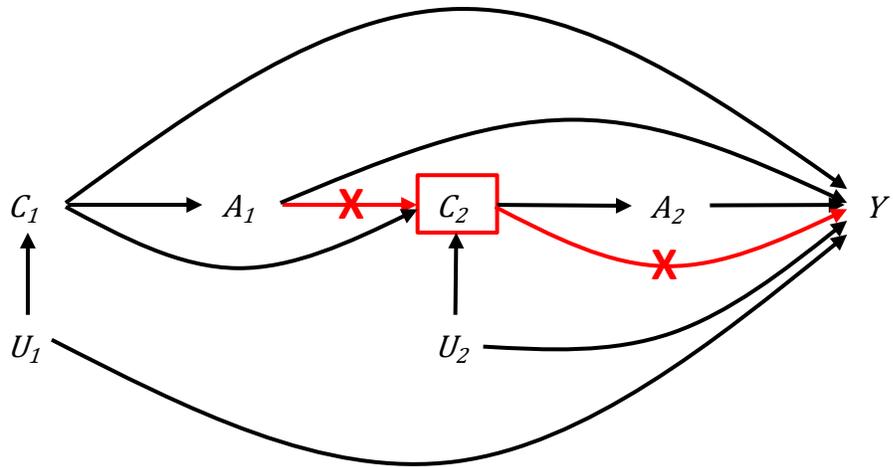

Figure 2. A stylized graph illustrating bias due to over-control of intermediate pathways

Notes: $A_t$ denotes a time-varying treatment, $C_t$ denotes an observed time-varying confounder, $U_t$ denotes an unobserved time-varying covariate, and $Y$ denotes an end-of-study outcome. A box around a variable denotes conditioning.

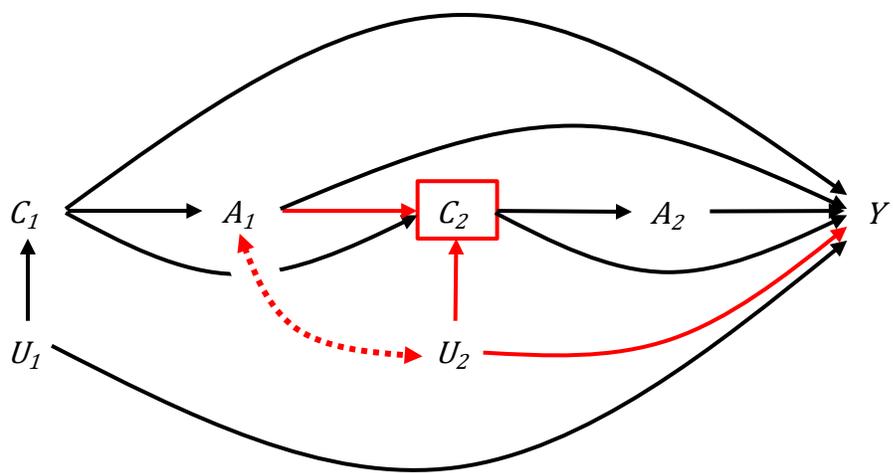

Figure 3. A stylized graph illustrating bias due to endogenous selection (or collider stratification)

Notes: $A_t$ denotes a time-varying treatment, $C_t$ denotes an observed time-varying confounder, $U_t$ denotes an unobserved time-varying covariate, and $Y$ denotes an end-of-study outcome. A box around a variable denotes conditioning. A bidirectional dashed line denotes a non-causal association between variables.

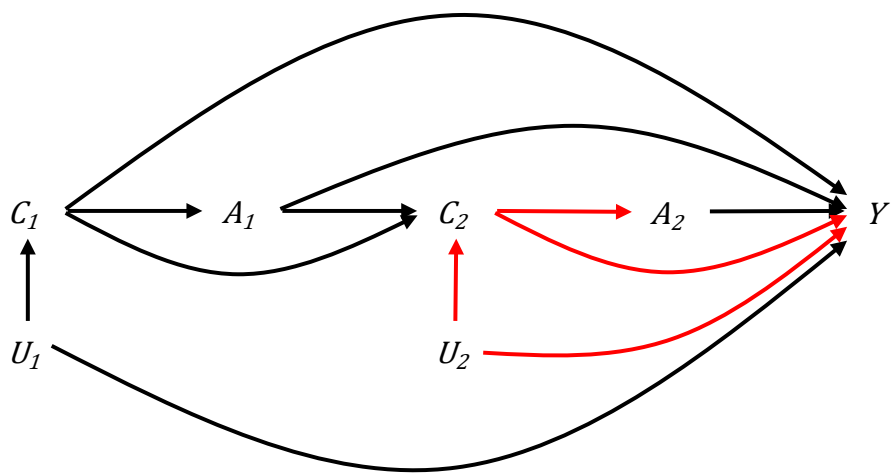

Figure 4. A stylized graph illustrating bias due to uncontrolled confounding

Notes: $A_t$ denotes a time-varying treatment, $C_t$ denotes an observed time-varying confounder, $U_t$ denotes an unobserved time-varying covariate, and $Y$ denotes an end-of-study outcome.

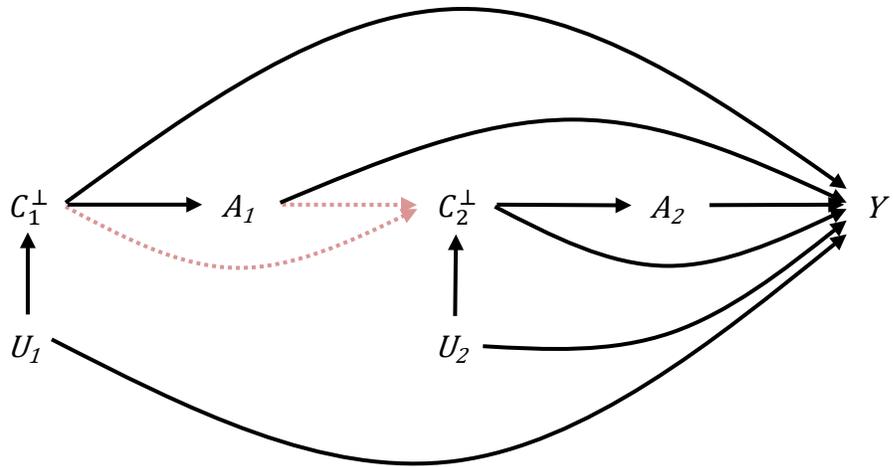

Figure 5. A stylized graph illustrating the logic of regression-with-residuals

Notes: $A_t$ denotes a time-varying treatment, $C_1^\perp = C_1 - E(C_1)$ and $C_2^\perp = C_2 - E(C_2|C_1, A_1)$ denote residualized time-varying confounders, $U_t$ denotes an unobserved time-varying covariate, and $Y$ denotes an end-of-study outcome.

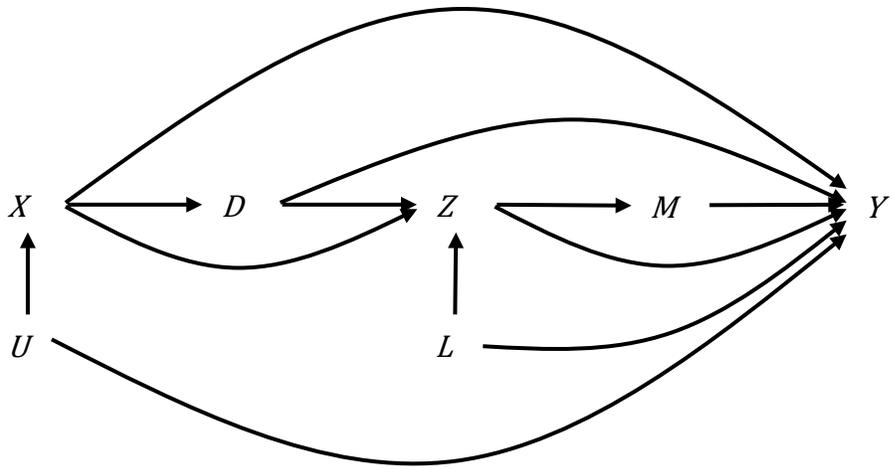

Figure 6. A directed acyclic graph illustrating a set of causal relationships between a treatment, a putative mediator, a set of confounders, and an outcome

Notes: $D$ denotes treatment, $M$ denotes the putative mediator, $X$ is a treatment-outcome confounder measured at baseline, $Z$ is a treatment-induced confounder of the mediator-outcome relationship, $U$ and $L$ denote unobserved covariates, and $Y$ denotes the outcome.

**TABLES**

Table 1. Performance of RWR relative to other estimators under different levels of treatment-outcome confounding

| Estimator/statistic | Magnitude of confounding | | | | |
|---|---|---|---|---|---|
| | $\gamma=0.1$ | $\gamma=0.2$ | $\gamma=0.3$ | $\gamma=0.4$ | $\gamma=0.5$ |
| Conventional regression | | | | | |
|   Bias | -0.150 | -0.200 | -0.252 | -0.299 | -0.351 |
|   SD | 0.134 | 0.137 | 0.141 | 0.145 | 0.151 |
|   RMSE | 0.201 | 0.242 | 0.288 | 0.332 | 0.382 |
| IPTW estimation | | | | | |
|   Bias | -0.001 | 0.002 | 0.010 | 0.035 | 0.085 |
|   SD | 0.135 | 0.147 | 0.176 | 0.228 | 0.296 |
|   RMSE | 0.135 | 0.147 | 0.176 | 0.230 | 0.308 |
| G-estimation | | | | | |
|   Bias | 0.000 | 0.000 | -0.001 | 0.002 | 0.000 |
|   SD | 0.134 | 0.139 | 0.145 | 0.152 | 0.163 |
|   RMSE | 0.134 | 0.139 | 0.145 | 0.152 | 0.163 |
| RWR w/o interactions | | | | | |
|   Bias | 0.000 | 0.000 | -0.001 | 0.002 | 0.000 |
|   SD | 0.134 | 0.139 | 0.145 | 0.151 | 0.161 |
|   RMSE | 0.134 | 0.139 | 0.145 | 0.151 | 0.161 |
| RWR w/ interactions | | | | | |
|   Bias | 0.000 | 0.000 | -0.001 | 0.002 | -0.001 |
|   SD | 0.134 | 0.140 | 0.146 | 0.154 | 0.164 |
|   RMSE | 0.134 | 0.140 | 0.146 | 0.154 | 0.164 |

Notes: SD denotes the standard deviation, and RMSE denotes the root mean squared error. Results are based on 10,000 simulations. See the Online Supplement for details.

Table 2. Performance of RWR relative to other estimators under different levels of effect moderation and a moderate-to-high level of confounding

| Estimator/statistic | Magnitude of effect moderation | | | | |
|---|---|---|---|---|---|
| | $\theta=0.1$ | $\theta=0.2$ | $\theta=0.3$ | $\theta=0.4$ | $\theta=0.5$ |
| Conventional regression | | | | | |
|    Bias | -0.369 | -0.439 | -0.508 | -0.575 | -0.645 |
|    SD | 0.145 | 0.149 | 0.151 | 0.155 | 0.163 |
|    RMSE | 0.396 | 0.463 | 0.530 | 0.595 | 0.665 |
| IPTW estimation | | | | | |
|    Bias | 0.022 | 0.024 | 0.023 | 0.028 | 0.021 |
|    SD | 0.235 | 0.246 | 0.261 | 0.274 | 0.299 |
|    RMSE | 0.236 | 0.247 | 0.262 | 0.275 | 0.300 |
| G-estimation | | | | | |
|    Bias | -0.023 | -0.047 | -0.071 | -0.094 | -0.119 |
|    SD | 0.155 | 0.161 | 0.164 | 0.168 | 0.177 |
|    RMSE | 0.157 | 0.168 | 0.179 | 0.193 | 0.213 |
| RWR w/o interactions | | | | | |
|    Bias | -0.037 | -0.076 | -0.115 | -0.151 | -0.192 |
|    SD | 0.154 | 0.161 | 0.166 | 0.171 | 0.182 |
|    RMSE | 0.159 | 0.178 | 0.202 | 0.228 | 0.264 |
| RWR w/ interactions | | | | | |
|    Bias | 0.001 | 0.001 | 0.001 | 0.000 | -0.001 |
|    SD | 0.156 | 0.161 | 0.164 | 0.167 | 0.175 |
|    RMSE | 0.156 | 0.161 | 0.164 | 0.167 | 0.175 |

Notes: SD denotes the standard deviation, and RMSE denotes the root mean squared error. Results are based on 10,000 simulations. See the Online Supplement for details.

Table 3. Estimated marginal effects of exposure to disadvantaged neighborhoods on end-of-study math achievement

| Estimator/statistic | $DTE\,(1,0)$ | | | $PTE\,(a_1,1)$ | | | $CTE$ | | |
|---|---|---|---|---|---|---|---|---|---|
| | Est | SE | | Est | SE | | Est | SE | |
| RWR with interactions | -0.034 | (0.049) | | -0.094 | (0.046) | * | -0.127 | (0.038) | *** |
| RWR without interactions | -0.030 | (0.044) | | -0.097 | (0.040) | * | -0.127 | (0.038) | *** |
| G-estimation | -0.032 | (0.040) | | -0.096 | (0.041) | * | -0.127 | (0.047) | ** |

Notes: Sample includes respondents who were interviewed at the 1997 wave of the CDS between age 3 and 7. Results are combined estimates from 100 imputations. The outcome is standardized to have zero mean and unit variance. SEs are based on the block bootstrap with 1,000 replications.

†$p < 0.10$, *$p < 0.05$, **$p < 0.01$, and ***$p < 0.001$ for two-sided tests of no effect.

Table 4. Estimated total and direct effects of college completion on depression

| Estimator/statistic | Total Effect | | | CDE (1,0.5) | |
|---|---|---|---|---|---|
| | Est | SE | | Est | SE |
| RWR with interactions | -0.165 | (0.066) | * | -0.103 | (0.066) |
| RWR without interactions | -0.089 | (0.053) | | -0.077 | (0.060) |
| G-estimation | -0.128 | (0.040) | * | -0.098 | (0.063) |

Notes: Sample includes respondents to the NLSY79 who were age 13-17 when first interviewed. SEs are based on the nonparametric bootstrap with 1,000 replications.

$\dagger p < 0.10$, $*p < 0.05$, $**p < 0.01$, and $***p < 0.001$ for two-sided tests of no effect.

**ONLINE SUPPLEMENT**

**Part A: R Code for Simulations**

```
###############################################
###############################################
#                                             #
# FILE NAME: 20_create_table_1.R              #
# PURPOSE: conduct simulation experiments     #
# NOTES: last edited by GW on 7/12/2018       #
#                                             #
###############################################
###############################################

set.seed(8675309)
nsim<-10000
simreg<-simiptw<-simgest<-simrwr<-simrwri<-matrix(data=NA,nrow=nsim,ncol=5)
obs<-500
for (k in seq(from=1,to=5,by=1)) {
        gamma<-(1/10)*k
        theta<-0
        for (i in 1:nsim) {
                ### SIMULATE DATA ###
                u<-rnorm(obs,0,1)
                c1<-rnorm(obs,0,1)
                a1<-rbinom(obs,1,pnorm(gamma*c1))
                c2<-rnorm(obs,0.5*u+0.5*c1+0.5*a1,1)
                a2<-rbinom(obs,1,pnorm(gamma*c1+0.5*a1+gamma*c2))
                y<-rnorm(obs,0.5*u+c1*gamma+a1*(0.2+c1*theta)+(c2-(0.5*c1+0.5*a1))*gamma+a2*(0.2+0.1*a1+(c1+(c2-
                (0.5*c1+0.5*a1)))*theta),1)
                ### CONVENTIONAL REGRESSION ###
                m1<-lm(y~c1+a1+c2+a2+a1*a2)
                simreg[i,k]<-(m1$coefficients[3]+m1$coefficients[5]+m1$coefficients[6])-0.5
```

```
rm(list=c('m1'))
### IPTW ###
m1<-glm(a1~1,family=binomial(link="probit"))
m2<-glm(a1~c1,family=binomial(link="probit"))
m3<-glm(a2~a1,family=binomial(link="probit"))
m4<-glm(a2~c1+a1+c2,family=binomial(link="probit"))
iptw1<-(m1$fitted.values/m2$fitted.values)*a1+((1-m1$fitted.values)/(1-m2$fitted.values))*(1-a1)
iptw2<-(m3$fitted.values/m4$fitted.values)*a2+((1-m3$fitted.values)/(1-m4$fitted.values))*(1-a2)
iptwf<-iptw1*iptw2
m5<-lm(y~a1+a2+a1*a2,weights=iptwf)
simiptw[i,k]<-(m5$coefficients[2]+m5$coefficients[3]+m5$coefficients[4])-0.5
rm(list=c('m1','m2','m3','m4','m5','iptw1','iptw2','iptwf'))
### G-ESTIMATION ###
ps1<-glm(a1~c1,family=binomial(link="probit"))$fitted.values
ps2<-glm(a2~c1+a1+c2,family=binomial(link="probit"))$fitted.values
m1<-lm(y~c1+ps1+a1+c2+ps2+a1*ps2+a2+a1*a2)
h<-y-a2*(m1$coefficients[7]+m1$coefficients[9]*a1)
m2<-lm(h~c1+ps1+a1)
simgest[i,k]<-(m2$coefficients[4]+m1$coefficients[7]+m1$coefficients[9])-0.5
rm(list=c('m1','m2','ps1','ps2','h'))
### RWR W/O INTERACTIONS ###
c1r<-lm(c1~1)$residuals
c2r<-lm(c2~c1+a1)$residuals
m1<-lm(y~c1r+a1+c2r+a2+a1*a2)
simrwr[i,k]<-(m1$coefficients[3]+m1$coefficients[5]+m1$coefficients[6])-0.5
rm(list=c('m1'))
### RWR W/ INTERACTIONS ###
m1<-lm(y~c1r+a1+c1r*a1+c2r+a2+a1*a2+c1r*a2+c2r*a2)
simrwri[i,k]<-(m1$coefficients[3]+m1$coefficients[5]+m1$coefficients[7])-0.5
rm(list=c('m1','c1r','c2r'))
}
}
```

```
sink("D:\\projects\\rwr_marginal_effects\\programs\\_LOGS\\20_create_table_1_log.txt")
cat("--------------------------------------------------------------------------------------------\n")
cat("Conventional Regression Estimates\n")
cat("--------------------------------------------------------------------------------------------\n")
summary(simreg)
cat("\n")
cat("Standard Deviations:\n")
for (j in 1:5) { print(sd(simreg[,j])) }
cat("--------------------------------------------------------------------------------------------\n")
cat("\n")
cat("--------------------------------------------------------------------------------------------\n")
cat("IPTW Estimates\n")
cat("--------------------------------------------------------------------------------------------\n")
summary(simiptw)
cat("\n")
cat("Standard Deviations:\n")
for (j in 1:5) { print(sd(simiptw[,j])) }
cat("--------------------------------------------------------------------------------------------\n")
cat("\n")
cat("--------------------------------------------------------------------------------------------\n")
cat("G-Estimates\n")
cat("--------------------------------------------------------------------------------------------\n")
summary(simgest)
cat("\n")
cat("Standard Deviations:\n")
for (j in 1:5) { print(sd(simgest[,j])) }
cat("--------------------------------------------------------------------------------------------\n")
cat("\n")
cat("--------------------------------------------------------------------------------------------\n")
cat("RWR (no interactions) Estimates\n")
cat("--------------------------------------------------------------------------------------------\n")
summary(simrwr)
cat("\n")
```

```
cat("Standard Deviations:\n")
for (j in 1:5) { print(sd(simrwr[,j])) }
cat("--------------------------------------------------------------------------------------------------\n")
cat("\n")
cat("--------------------------------------------------------------------------------------------------\n")
cat("RWR (with interactions) Estimates\n")
cat("--------------------------------------------------------------------------------------------------\n")
summary(simrwri)
cat("\n")
cat("Standard Deviations:\n")
for (j in 1:5) { print(sd(simrwri[,j])) }
cat("--------------------------------------------------------------------------------------------------\n")
sink()

################################################
################################################
#                                              #
# FILE NAME: 21_create_table_2.R               #
# PURPOSE: conduct simulation experiments       #
# NOTES: last edited by GW on 7/12/2018        #
#                                              #
################################################
################################################

set.seed(90210)
nsim<-10000
simreg<-simiptw<-simgest<-simrwr<-simrwri<-matrix(data=NA,nrow=nsim,ncol=5)
obs<-500
for (k in seq(from=1,to=5,by=1)) {
        gamma<--0.4
        theta<--(1/10)*k
        for (i in 1:nsim) {
                ### SIMULATE DATA ###
```

```
u<-rnorm(obs,0,1)
c1<-rnorm(obs,0,1)
a1<-rbinom(obs,1,pnorm(gamma*c1))
c2<-rnorm(obs,0.5*u+0.5*c1+0.5*a1,1)
a2<-rbinom(obs,1,pnorm(gamma*c1+0.5*a1+gamma*c2))
y<-rnorm(obs,0.5*u+c1*gamma+a1*(0.2+c1*theta)+(c2-(0.5*c1+0.5*a1))*gamma+a2*(0.2+0.1*a1+(c1+(c2-
(0.5*c1+0.5*a1)))*theta),1)
### CONVENTIONAL REGRESSION ###
m1<-lm(y~c1+a1+c2+a2+a1*a2)
simreg[i,k]<-(m1$coefficients[3]+m1$coefficients[5]+m1$coefficients[6])-0.5
rm(list=c('m1'))
### IPTW ###
m1<-glm(a1~1,family=binomial(link="probit"))
m2<-glm(a1~c1,family=binomial(link="probit"))
m3<-glm(a2~a1,family=binomial(link="probit"))
m4<-glm(a2~c1+a1+c2,family=binomial(link="probit"))
iptw1<-(m1$fitted.values/m2$fitted.values)*a1+((1-m1$fitted.values)/(1-m2$fitted.values))*(1-a1)
iptw2<-(m3$fitted.values/m4$fitted.values)*a2+((1-m3$fitted.values)/(1-m4$fitted.values))*(1-a2)
iptwf<-iptw1*iptw2
m5<-lm(y~a1+a2+a1*a2,weights=iptwf)
simiptw[i,k]<-(m5$coefficients[2]+m5$coefficients[3]+m5$coefficients[4])-0.5
rm(list=c('m1','m2','m3','m4','m5','iptw1','iptw2','iptwf'))
### G-ESTIMATION ###
ps1<-glm(a1~c1,family=binomial(link="probit"))$fitted.values
ps2<-glm(a2~c1+a1+c2,family=binomial(link="probit"))$fitted.values
m1<-lm(y~c1+ps1+a1+c2+ps2+a1*ps2+a2+a1*a2)
h<-y-a2*(m1$coefficients[7]+m1$coefficients[9]*a1)
m2<-lm(h~c1+ps1+a1)
simgest[i,k]<-(m2$coefficients[4]+m1$coefficients[7]+m1$coefficients[9])-0.5
rm(list=c('m1','m2','ps1','ps2','h'))
### RWR W/O INTERACTIONS ###
c1r<-lm(c1~1)$residuals
c2r<-lm(c2~c1+a1)$residuals
```

```
            m1<-lm(y~c1r+a1+c2r+a2+a1*a2)
            simrwr[i,k]<-(m1$coefficients[3]+m1$coefficients[5]+m1$coefficients[6])-0.5
            rm(list=c('m1'))
            ### RWR W/ INTERACTIONS ###
            m1<-lm(y~c1r+a1+c1r*a1+c2r+a2+a1*a2+c1r*a2+c2r*a2)
            simrwri[i,k]<-(m1$coefficients[3]+m1$coefficients[5]+m1$coefficients[7])-0.5
            rm(list=c('m1','c1r','c2r'))
            }
      }

sink("D:\\projects\\rwr_marginal_effects\\programs\\_LOGS\\21_create_table_2_log.txt")
cat("----------------------------------------------------------------------------------------------\n")
cat("Conventional Regression Estimates\n")
cat("----------------------------------------------------------------------------------------------\n")
summary(simreg)
cat("\n")
cat("Standard Deviations:\n")
for (j in 1:5) { print(sd(simreg[,j])) }
cat("----------------------------------------------------------------------------------------------\n")
cat("\n")
cat("----------------------------------------------------------------------------------------------\n")
cat("IPTW Estimates\n")
cat("----------------------------------------------------------------------------------------------\n")
summary(simiptw)
cat("\n")
cat("Standard Deviations:\n")
for (j in 1:5) { print(sd(simiptw[,j])) }
cat("----------------------------------------------------------------------------------------------\n")
cat("\n")
cat("----------------------------------------------------------------------------------------------\n")
cat("G-Estimates\n")
cat("----------------------------------------------------------------------------------------------\n")
summary(simgest)
```

```
cat("\n")
cat("Standard Deviations:\n")
for (j in 1:5) { print(sd(simgest[,j])) }
cat("----------------------------------------------------------------------------------------\n")
cat("\n")
cat("----------------------------------------------------------------------------------------\n")
cat("RWR (no interactions) Estimates\n")
cat("----------------------------------------------------------------------------------------\n")
summary(simrwr)
cat("\n")
cat("Standard Deviations:\n")
for (j in 1:5) { print(sd(simrwr[,j])) }
cat("----------------------------------------------------------------------------------------\n")
cat("\n")
cat("----------------------------------------------------------------------------------------\n")
cat("RWR (with interactions) Estimates\n")
cat("----------------------------------------------------------------------------------------\n")
summary(simrwri)
cat("\n")
cat("Standard Deviations:\n")
for (j in 1:5) { print(sd(simrwri[,j])) }
cat("----------------------------------------------------------------------------------------\n")
sink()
```

**Part B: R Code for Analyses of Neighborhood Effects on Math Achievement**

```
################################################
################################################
##                            ##
## PROGRAM NAME: 22_create_table_3        ##
## AUTHOR: GW                 ##
## DATE: 7/12/2018             ##
## DESCRIPTION:               ##
##                      ##
## computes marginal effect estimates from  ##
## PSID-CDS using RWR w/o interactions,     ##
## RWR w/ all two-way treatment x cov      ##
## interactions, and g-estimation; computes ##
## block boostrap standard errors         ##
##                       ##
################################################
################################################

rm(list=ls())
library(foreign)
library(dplyr)
library(tidyr)
library(CBPS)
library(ggplot2)
library(mgcv)
nmi<-100
nboot<-1000

###################
#INPUT/RECODE DATA#
###################
psidmi<-read.dta("U:\\rwr_marginal_effects\\data\\v03_psid_merged_nvars_mi.dta")
```

```
vars<-
c("ncaseid","nfamid97","ncohort","nblack","nfemale","npcgeduc","nincneedt0","nhdmarstatt0","napscoret0","nhadvgt1","napscoret1"
,"nincneedt1","nhdmarstatt1","nhadvgt2","napscoret2","_mi","_mj")
psidmi<-psidmi[which(psidmi$"_mj"!=0),vars]
psidmi$nhdadvgt1<-((psidmi$nhadvgt1-mean(psidmi$nhadvgt1))/sd(psidmi$nhadvgt1))*(-1)
psidmi$nhdadvgt2<-(psidmi$nhadvgt2-mean(psidmi$nhadvgt2))/sd(psidmi$nhadvgt2)*(-1)
psidmi$napscoret0<-(psidmi$napscoret0-mean(psidmi$napscoret0))/sd(psidmi$napscoret0)
psidmi$napscoret1<-(psidmi$napscoret1-mean(psidmi$napscoret1))/sd(psidmi$napscoret1)
psidmi$napscoret2<-(psidmi$napscoret2-mean(psidmi$napscoret2))/sd(psidmi$napscoret2)
psidmi$nfemale<-as.numeric(psidmi$nfemale)-1

###############################################################
#COMPUTE RWR ESTIMATES W/ ALL TWO-WAY TREATMENT X COV INTERACTIONS#
###############################################################
mibeta<-matrix(data=NA,nrow=nmi,ncol=3)
mivar<-matrix(data=NA,nrow=nmi,ncol=3)
for (i in 1:nmi) {
        # COMPUTE POINT ESTIMATES #
        psid<-psidmi[which(psidmi$"_mj"==i),]
        residualizet0<-function(y) { residuals(lm(y~1,data=psid)) }
        residualizet1<-function(y) {
        residuals(lm(y~ncohort+nblack+nfemale+npcgeduc+napscoret0+nincneedt0+nhdmarstatt0+nhdadvgt1,data=psid)) }
        psid<-psid %>% mutate_at(vars(ncohort,nblack,nfemale,npcgeduc,napscoret0,nincneedt0,nhdmarstatt0),
        funs(res=residualizet0))
        psid<-psid %>% mutate_at(vars(nincneedt1,nhdmarstatt1,napscoret1), funs(res=residualizet1))
        rwr<-lm(napscoret2~nhdadvgt1+nhdadvgt2+ncohort_res+nblack_res+nfemale_res+npcgeduc_res+napscoret0_res+
        nincneedt0_res+nhdmarstatt0_res+napscoret1_res+nincneedt1_res+nhdmarstatt1_res+
        ncohort_res*nhdadvgt1+nblack_res*nhdadvgt1+nfemale_res*nhdadvgt1+npcgeduc_res*nhdadvgt1+napscoret0_res*nhdadv1
        +nincneedt0_res*nhdadvgt1+nhdmarstatt0_res*nhdadvgt1+ncohort_res*nhdadvgt2+nblack_res*nhdadvgt2+nfemale_res*nhd
        advgt2+npcgeduc_res*nhdadvgt2+napscoret0_res*nhdadvgt2+nincneedt0_res*nhdadvgt2+nhdmarstatt0_res*nhdadvgt2+
        napscoret1_res*nhdadvgt2+nincneedt1_res*nhdadvgt2+nhdmarstatt1_res*nhdadvgt2,data=psid)
        mibeta[i,1]<-rwr$coef[2]
        mibeta[i,2]<-rwr$coef[3]
```

```
mibeta[i,3]<-rwr$coef[2]+rwr$coef[3]
# COMPUTE BLOCK BOOTSTRAP SEs #
set.seed(8675309)
bootdist<-matrix(data=NA,nrow=nboot,ncol=3)
for (j in 1:nboot) {
        idboot.1<-sample(unique(psid$nfamid97),replace=T)
        idboot.2<-table(idboot.1)
        psid.boot<-NULL
        for (k in 1:max(idboot.2)) {
                boot.data<-psid[psid$nfamid97 %in% names(idboot.2[idboot.2 %in% k]),]
                for (l in 1:k) { psid.boot<-rbind(psid.boot,boot.data) }
                }
        residualizet0<-function(y) { residuals(lm(y~1,data=psid.boot)) }
        residualizet1<-function(y) {
        residuals(lm(y~ncohort+nblack+nfemale+npcgeduc+napscoret0+nincneedt0+nhdmarstatt0+nhdadvgt1,data=psid.boot)
        ) }
        psid.boot<-psid.boot %>% mutate_at(vars(ncohort,nblack,nfemale,npcgeduc,napscoret0,nincneedt0,nhdmarstatt0),
        funs(res=residualizet0))
        psid.boot<-psid.boot %>% mutate_at(vars(nincneedt1,nhdmarstatt1,napscoret1), funs(res=residualizet1))
        rwrboot<-
        lm(napscoret2~nhdadvgt1+nhdadvgt2+ncohort_res+nblack_res+nfemale_res+npcgeduc_res+napscoret0_res+nincneedt
        0_res+nhdmarstatt0_res+napscoret1_res+nincneedt1_res+nhdmarstatt1_res+ncohort_res*nhdadvgt1+nblack_res*nhda
        dvgt1+nfemale_res*nhdadvgt1+npcgeduc_res*nhdadvgt1+napscoret0_res*nhdadvgt1+nincneedt0_res*nhdadvgt1+nhd
        marstatt0_res*nhdadvgt1+ncohort_res*nhdadvgt2+nblack_res*nhdadvgt2+nfemale_res*nhdadvgt2+npcgeduc_res*nhd
        advgt2+napscoret0_res*nhdadvgt2+nincneedt0_res*nhdadvgt2+nhdmarstatt0_res*nhdadvgt2+napscoret1_res*nhdadv
        gt2+nincneedt1_res*nhdadvgt2+nhdmarstatt1_res*nhdadvgt2,data=psid.boot)
        bootdist[j,1]<-rwrboot$coef[2]
        bootdist[j,2]<-rwrboot$coef[3]
        bootdist[j,3]<-rwrboot$coef[2]+rwrboot$coef[3]
        }
for (m in 1:3) { mivar[i,m]<-var(bootdist[,m]) }
        }
# COMBINE MI ESTIMATES #
```

```r
rwrwiest<-matrix(data=NA,nrow=3,ncol=4)
for (i in 1:3) {
        rwrwiest[i,1]<-round(mean(mibeta[,i]),digits=3)
        rwrwiest[i,2]<-round(sqrt(mean(mivar[,i])+(var(mibeta[,i])*(1+(1/nmi)))),digits=3)
        rwrwiest[i,3]<-round((rwrwiest[i,1]/rwrwiest[i,2]),digits=3)
        rwrwiest[i,4]<-round((pnorm(abs(rwrwiest[i,3]),0,1,lower.tail=FALSE)*2),digits=3)
        }

#######################################
#COMPUTE RWR ESTIMATES W/O INTERACTIONS#
#######################################
mibeta<-matrix(data=NA,nrow=nmi,ncol=3)
mivar<-matrix(data=NA,nrow=nmi,ncol=3)
for (i in 1:nmi) {
        # COMPUTE POINT ESTIMATES #
        psid<-psidmi[which(psidmi$"_mj"==i),]
        residualizet0<-function(y) { residuals(lm(y~1,data=psid)) }
        residualizet1<-function(y) {
        residuals(lm(y~ncohort+nblack+nfemale+npcgeduc+napscoret0+nincneedt0+nhdmarstatt0+nhdadvgt1,data=psid)) }
        psid<-psid %>% mutate_at(vars(ncohort,nblack,nfemale,npcgeduc,napscoret0,nincneedt0,nhdmarstatt0),
        funs(res=residualizet0))
        psid<-psid %>% mutate_at(vars(nincneedt1,nhdmarstatt1,napscoret1), funs(res=residualizet1))
        rwr<-
        lm(napscoret2~nhdadvgt1+nhdadvgt2+ncohort_res+nblack_res+nfemale_res+npcgeduc_res+napscoret0_res+nincneedt0_res+
        nhdmarstatt0_res+napscoret1_res+nincneedt1_res+nhdmarstatt1_res,data=psid)
        mibeta[i,1]<-rwr$coef[2]
        mibeta[i,2]<-rwr$coef[3]
        mibeta[i,3]<-rwr$coef[2]+rwr$coef[3]
        # COMPUTE BLOCK BOOTSTRAP SEs #
        set.seed(8675309)
        bootdist<-matrix(data=NA,nrow=nboot,ncol=3)
        for (j in 1:nboot) {
                idboot.1<-sample(unique(psid$nfamid97),replace=T)
```

```
                 idboot.2<-table(idboot.1)
                 psid.boot<-NULL
                 for (k in 1:max(idboot.2)) {
                          boot.data<-psid[psid$nfamid97 %in% names(idboot.2[idboot.2 %in% k]),]
                          for (l in 1:k) { psid.boot<-rbind(psid.boot,boot.data) }
                          }
                 residualizet0<-function(y) { residuals(lm(y~1,data=psid.boot)) }
                 residualizet1<-function(y) {
                 residuals(lm(y~ncohort+nblack+nfemale+npcgeduc+napscoret0+nincneedt0+nhdmarstatt0+nhdadvgt1,data=psid.boot)
                 ) }
                 psid.boot<-psid.boot %>% mutate_at(vars(ncohort,nblack,nfemale,npcgeduc,napscoret0,nincneedt0,nhdmarstatt0),
                 funs(res=residualizet0))
                 psid.boot<-psid.boot %>% mutate_at(vars(nincneedt1,nhdmarstatt1,napscoret1), funs(res=residualizet1))
                 rwrboot<-
                 lm(napscoret2~nhdadvgt1+nhdadvgt2+ncohort_res+nblack_res+nfemale_res+npcgeduc_res+napscoret0_res+nincneedt
                 0_res+nhdmarstatt0_res+napscoret1_res+nincneedt1_res+nhdmarstatt1_res,data=psid.boot)
                 bootdist[j,1]<-rwrboot$coef[2]
                 bootdist[j,2]<-rwrboot$coef[3]
                 bootdist[j,3]<-rwrboot$coef[2]+rwrboot$coef[3]
                 }
         for (m in 1:3) { mivar[i,m]<-var(bootdist[,m]) }
         }
# COMBINE MI ESTIMATES #
rwrest<-matrix(data=NA,nrow=3,ncol=4)
for (i in 1:3) {
         rwrest[i,1]<-round(mean(mibeta[,i]),digits=3)
         rwrest[i,2]<-round(sqrt(mean(mivar[,i])+(var(mibeta[,i])*(1+(1/nmi)))),digits=3)
         rwrest[i,3]<-round((rwrest[i,1]/rwrest[i,2]),digits=3)
         rwrest[i,4]<-round((pnorm(abs(rwrest[i,3]),0,1,lower.tail=FALSE)*2),digits=3)
         }

######################
# COMPUTE G-ESTIMATES #
```

```
#######################
mibeta<-matrix(data=NA,nrow=nmi,ncol=3)
mivar<-matrix(data=NA,nrow=nmi,ncol=3)
for (i in 1:nmi) {
        # COMPUTE POINT ESTIMATES #
        psid<-psidmi[which(psidmi$"_mj"==i),]
        psm1<-gam(nhdadvgt1~te(ncohort)+nblack+nfemale+te(npcgeduc)+te(napscoret0)+te(nincneedt0)+nhdmarstatt0,data=psid)
        psmfit1<-predict(psm1,type="response",se=T)
        ps1<-psmfit1$fit
        psm2<-
        gam(nhdadvgt2~te(ncohort)+nblack+nfemale+te(npcgeduc)+te(napscoret0)+te(nincneedt0)+nhdmarstatt0+te(nhdadvgt1)+te(n
        apscoret1)+te(nincneedt1)+nhdmarstatt1,data=psid)
        psmfit2<-predict(psm2,type="response",se=T)
        ps2<-psmfit2$fit
        m1<-
        lm(napscoret2~nhdadvgt1+nhdadvgt2+ncohort+nblack+nfemale+npcgeduc+napscoret0+nincneedt0+nhdmarstatt0+ps1+napsc
        oret1+nincneedt1+nhdmarstatt1+ps2,data=psid)
        psid$h<-psid$napscoret2-psid$nhdadvgt2*m1$coef[3]
        m2<-lm(h~nhdadvgt1+ncohort+nblack+nfemale+npcgeduc+napscoret0+nincneedt0+nhdmarstatt0+ps1,data=psid)
        mibeta[i,1]<-m2$coef[2]
        mibeta[i,2]<-m1$coef[3]
        mibeta[i,3]<-m2$coef[2]+m1$coef[3]
        # COMPUTE BLOCK BOOTSTRAP SEs #
        set.seed(8675309)
        bootdist<-matrix(data=NA,nrow=nboot,ncol=3)
        for (j in 1:nboot) {
                idboot.1<-sample(unique(psid$nfamid97),replace=T)
                idboot.2<-table(idboot.1)
                psid.boot<-NULL
                for (k in 1:max(idboot.2)) {
                        boot.data<-psid[psid$nfamid97 %in% names(idboot.2[idboot.2 %in% k]),]
                        for (l in 1:k) { psid.boot<-rbind(psid.boot,boot.data) }
                        }
```

```
                    psm1<-
                    gam(nhdadvgt1~te(ncohort)+nblack+nfemale+te(npcgeduc)+te(napscoret0)+te(nincneedt0)+nhdmarstatt0,data=psid.bo
                    ot)
                    psmfit1<-predict(psm1,type="response",se=T)
                    ps1<-psmfit1$fit
                    psm2<-
                    gam(nhdadvgt2~te(ncohort)+nblack+nfemale+te(npcgeduc)+te(napscoret0)+te(nincneedt0)+nhdmarstatt0+te(nhdadvgt
                    1)+te(napscoret1)+te(nincneedt1)+nhdmarstatt1,data=psid.boot)
                    psmfit2<-predict(psm2,type="response",se=T)
                    ps2<-psmfit2$fit
                    m1boot<-
                    lm(napscoret2~nhdadvgt1+nhdadvgt2+ncohort+nblack+nfemale+npcgeduc+napscoret0+nincneedt0+nhdmarstatt0+ps1
                    +napscoret1+nincneedt1+nhdmarstatt1+ps2,data=psid.boot)
                    psid.boot$h<-psid.boot$napscoret2-psid.boot$nhdadvgt2*m1$coef[3]
                    m2boot<-
                    lm(h~nhdadvgt1+ncohort+nblack+nfemale+npcgeduc+napscoret0+nincneedt0+nhdmarstatt0+ps1,data=psid.boot)
                    bootdist[j,1]<-m2boot$coef[2]
                    bootdist[j,2]<-m1boot$coef[3]
                    bootdist[j,3]<-m2boot$coef[2]+m1boot$coef[3]
                    }
          for (m in 1:3) { mivar[i,m]<-var(bootdist[,m]) }
          }
# COMBINE MI ESTIMATES #
gest<-matrix(data=NA,nrow=3,ncol=4)
for (i in 1:3) {
          gest[i,1]<-round(mean(mibeta[,i]),digits=3)
          gest[i,2]<-round(sqrt(mean(mivar[,i])+(var(mibeta[,i])*(1+(1/nmi)))),digits=3)
          gest[i,3]<-round((gest[i,1]/gest[i,2]),digits=3)
          gest[i,4]<-round((pnorm(abs(gest[i,3]),0,1,lower.tail=FALSE)*2),digits=3)
          }

##############
#PRINT RESULTS#
```

```
###############
sink("U:\\rwr_marginal_effects\\programs\\_LOGS\\22_create_table_3_log.txt")
cat("~~~~~~~~~~~~~~~~~~~~~~~~~~~~~~~~~~~~~~~~~~~~~\n")
cat("~~~~~~~~~~~~~~~~~~~~~~~~~~~~~~~~~~~~~~~~~~~~~\n")
cat("=========================================\n")
cat("RWR w/ All Two-way A x C Interactions\n")
cat("=========================================\n")
print(rwrwiest)
cat("=========================================\n")
cat("RWR w/o Interactions\n")
cat("=========================================\n")
print(rwrest)
cat("=========================================\n")
cat("G-Estimation\n")
cat("=========================================\n")
print(gest)
cat("~~~~~~~~~~~~~~~~~~~~~~~~~~~~~~~~~~~~~~~~~~~~~\n")
cat("~~~~~~~~~~~~~~~~~~~~~~~~~~~~~~~~~~~~~~~~~~~~~\n")
cat("Note: Table Columns = Est / SE / Z / Pvalue\n")
sink()
```

**Part C: R Code for Analyses of Education Effects on Depression**

```
#################################################
#################################################
##                              ##
## PROGRAM NAME: 23_create_table_4        ##
## AUTHOR: XZ                   ##
## DATE: 7/12/2018             ##
## DESCRIPTION:                ##
##                      ##
## computes marginal effect estimates from  ##
## NLSY using RWR w/o interactions,      ##
## RWR w/ all two-way treatment x cov     ##
## interactions, and g-estimation; computes ##
## boostrap standard errors         ##
##                      ##
#################################################
#################################################

rm(list=ls(all=TRUE))

library("haven")
library("Hmisc")
library("readr")
library("tidyr")
library("dplyr")
library("pryr")
library("survey")

load("nlsy79_dpr.RData")

# g estimation
```

```
nlsy79_dpr$ps_college <- glm(college ~ male + black + test_score + educ_exp +
              father + hispanic + urban + educ_mom + num_sibs,
         family = binomial("probit"), weights = weights, data = nlsy79_dpr)$fitted.values

nlsy79_dpr$ps_ses <- lm(ses ~ male + black + test_score + educ_exp +
         father + hispanic + urban + educ_mom + num_sibs +
         cesd92 + prmarr98 + transitions98, weights = weights, data = nlsy79_dpr)$fitted.values

m1 <- lm(cesd40 ~ college + male + black + test_score + educ_exp + father + hispanic +
        urban + educ_mom + num_sibs + cesd92 + prmarr98 + transitions98 + college * ses + college * ps_ses,
      weights = weights, data = nlsy79_dpr)

nlsy79_dpr$cesd40_demed <-  nlsy79_dpr$cesd40 - nlsy79_dpr$ses * m1$coef["ses"] -
  nlsy79_dpr$college * nlsy79_dpr$ses * m1$coef["college:ses"]

m2 <- lm(cesd40_demed ~ college + male + black + test_score + educ_exp + father + hispanic + urban + educ_mom + num_sibs +
ps_college,
      weights = weights, data = nlsy79_dpr)

m0 <- lm(cesd40 ~ college + male + black + test_score + educ_exp + father + hispanic + urban + educ_mom + num_sibs +
ps_college,
      weights = weights, data = nlsy79_dpr)

# RWR

residualize <- function(y, df) {
  residuals(lm(y ~ college + male + black + test_score + educ_exp +
             father + hispanic + urban + educ_mom + num_sibs,
             weights = weights, data = df, na.action = na.exclude))
}

nlsy79_dpr_rwr <- nlsy79_dpr  %>%
  mutate_at(vars(cesd92:transitions98), funs(res = residualize(., df = nlsy79_dpr)))
```

```
# RWR without effect modifications

overall <- lm(cesd40 ~ college +
            male + black + test_score + educ_exp + father +
            hispanic + urban + educ_mom + num_sibs,
         weights = weights, data = nlsy79_dpr_rwr)

rwr <- lm(cesd40 ~ college +
         male + black + test_score + educ_exp + father +
         hispanic + urban + educ_mom + num_sibs +
         ses + college * ses +
         cesd92_res + prmarr98_res + transitions98_res,
       weights = weights, data = nlsy79_dpr_rwr)

summary(rwr)

# RWR with effect modifications

overall_effmod <- lm(cesd40 ~ college +
               (male + black + test_score + educ_exp + father +
                 hispanic + urban + educ_mom + num_sibs) * college,
             weights = weights, data = nlsy79_dpr_rwr)

rwr_effmod <- lm(cesd40 ~ college +
               (male + black + test_score + educ_exp + father +
                 hispanic + urban + educ_mom + num_sibs) * college +
               ses + college * ses +
               (male + black + test_score + educ_exp + father +
                 hispanic + urban + educ_mom + num_sibs +
                 cesd92_res + prmarr98_res + transitions98_res) * ses,
             weights = weights, data = nlsy79_dpr_rwr)
```

```
# bootstrap RWR

K <- 1000

ate_g_boot <- cde_g_ses_boot <- rep(NA, K)
ate_boot <- cde_ses_boot <- rep(NA, K)
ate_effmod_boot <- cde_effmod_ses_boot <- rep(NA, K)

for (k in seq(1, K)){

  cat(k, "\n")

  nlsy79_dpr_boot <- nlsy79_dpr[sample(nrow(nlsy79_dpr), replace = TRUE), ]

  # g boot

  nlsy79_dpr_boot$ps_college <- glm(college ~ male + black + test_score + educ_exp +
                        father + hispanic + urban + educ_mom + num_sibs,
                      family = binomial("probit"), weights = weights, data = nlsy79_dpr_boot)$fitted.values

  nlsy79_dpr_boot$ps_ses <- lm(ses ~ male + black + test_score + educ_exp +
                        father + hispanic + urban + educ_mom + num_sibs +
                        cesd92 + prmarr98 + transitions98, weights = weights, data = nlsy79_dpr_boot)$fitted.values

  m1_boot <- lm(cesd40 ~ college + male + black + test_score + educ_exp + father + hispanic +
                urban + educ_mom + num_sibs + cesd92 + prmarr98 + transitions98 + college * ses + college * ps_ses,
          weights = weights, data = nlsy79_dpr_boot)

  nlsy79_dpr_boot$cesd40_demed <-  nlsy79_dpr_boot$cesd40 - nlsy79_dpr_boot$ses * m1_boot$coef["ses"] -
    nlsy79_dpr_boot$college * nlsy79_dpr_boot$ses * m1_boot$coef["college:ses"]

  m2_boot <- lm(cesd40_demed ~ college + male + black + test_score + educ_exp + father + hispanic + urban + educ_mom +
  num_sibs + ps_college,
```

```
          weights = weights, data = nlsy79_dpr_boot)

 m0_boot <- lm(cesd40 ~ college + male + black + test_score + educ_exp + father + hispanic + urban + educ_mom + num_sibs +
ps_college,
          weights = weights, data = nlsy79_dpr_boot)

 # RWR boot

 nlsy79_dpr_rwr_boot <- nlsy79_dpr_boot  %>%
   mutate_at(vars(cesd92:transitions98), funs(res = residualize(., df = nlsy79_dpr_boot)))

 overall_boot <- lm(cesd40 ~ college +
                   male + black + test_score + educ_exp + father +
                   hispanic + urban + educ_mom + num_sibs,
                weights = weights, data = nlsy79_dpr_rwr_boot)

 rwr_boot <- lm(cesd40 ~ college +
              male + black + test_score + educ_exp + father +
              hispanic + urban + educ_mom + num_sibs +
              ses + college * ses +
              cesd92_res + prmarr98_res + transitions98_res,
            weights = weights, data = nlsy79_dpr_rwr_boot)

 overall_effmod_boot <- lm(cesd40 ~ college +
                      (male + black + test_score + educ_exp + father +
                        hispanic + urban + educ_mom + num_sibs) * college,
                   weights = weights, data = nlsy79_dpr_rwr_boot)

 rwr_effmod_boot <- lm(cesd40 ~ college +
                   (male + black + test_score + educ_exp + father +
                     hispanic + urban + educ_mom + num_sibs) * college +
                   ses + college * ses +
                   (male + black + test_score + educ_exp + father +
```

```
                    hispanic + urban + educ_mom + num_sibs +
                    cesd92_res + prmarr98_res + transitions98_res) * ses,
                weights = weights, data = nlsy79_dpr_rwr_boot)

    ate_g_boot[k] <- coef(summary(m0_boot))["college", 1]
    cde_g_ses_boot[k] <- coef(summary(m2_boot))["college", 1]

    ate_boot[k] <- coef(summary(overall_boot))["college", 1]
    cde_ses_boot[k] <- coef(summary(rwr_boot))["college", 1]

    ate_effmod_boot[k] <- coef(summary(overall_effmod_boot))["college", 1]
    cde_effmod_ses_boot[k] <- coef(summary(rwr_effmod_boot))["college", 1]

}

# output

ate_g_est <- coef(summary(m0))["college", 1]
cde_g_ses_est <- coef(summary(m2))["college", 1]

ate_g_se <- sd(ate_g_boot)
cde_g_ses_se <- sd(cde_g_ses_boot)

ate_g_p <- 2 * mean(ate_g_boot>0)
cde_g_ses_p <- 2 * mean(cde_g_ses_boot>0)

ate_est <- coef(summary(overall))["college", 1]
cde_ses_est <- coef(summary(rwr))["college", 1]

ate_se <- sd(ate_boot)
cde_ses_se <- sd(cde_ses_boot)

ate_p <- 2 * mean(ate_boot>0)
```

```
cde_ses_p <- 2 * mean(cde_ses_boot>0)

ate_effmod_est <- coef(summary(overall_effmod))["college", 1]
cde_effmod_ses_est <- coef(summary(rwr_effmod))["college", 1]

ate_effmod_se <- sd(ate_effmod_boot)
cde_effmod_ses_se <- sd(cde_effmod_ses_boot)

ate_effmod_p <- 2 * mean(ate_effmod_boot>0)
cde_effmod_ses_p <- 2 * mean(cde_effmod_ses_boot>0)

ate_g <- c(ate_g_est, ate_g_se, ate_g_p)
cde_g_ses <- c(cde_g_ses_est, cde_g_ses_se, cde_g_ses_p)

ate <- c(ate_est, ate_se, ate_p)
cde_ses <- c(cde_ses_est, cde_ses_se, cde_ses_p)

ate_effmod <- c(ate_effmod_est, ate_effmod_se, ate_effmod_p)
cde_effmod_ses <- c(cde_effmod_ses_est, cde_effmod_ses_se, cde_effmod_ses_p)

out <- rbind(c(ate_effmod, cde_effmod_ses), c(ate, cde_ses), c(ate_g, cde_g_ses))

write.csv(out, "23_create_table4.csv", row.names = FALSE)
```